\documentclass[11pt]{article}
\usepackage{amsfonts}
\usepackage{amssymb}
\usepackage{graphicx}
\usepackage{amsmath}
\usepackage{amsthm}
\usepackage{color}
\usepackage{multirow}
\usepackage{url}
\usepackage{float}
\usepackage{booktabs} 
\usepackage{makecell}
\usepackage{listings}
\usepackage{subcaption}
\usepackage{algorithm}
\usepackage{enumerate}
\usepackage{tikz}
\usepackage{algpseudocode}
\lstnewenvironment{Rcode}{\lstset{language=R}}{}
\newtheorem{theorem}{Theorem}[section]

\newtheorem{lemma}[theorem]{Lemma}

\definecolor{battleshipgrey}{rgb}{0.52, 0.52, 0.51}
\definecolor{navyblue}{rgb}{0.0, 0.0, 0.5}
\definecolor{arsenic}{rgb}{0.23, 0.27, 0.29}
\definecolor{oldmauve}{rgb}{0.4, 0.19, 0.28}
\usepackage[colorlinks=TRUE]{hyperref}
\hypersetup{
	colorlinks=true,
	linkcolor=navyblue,
	filecolor=blue,      
	urlcolor=oldmauve,
	citecolor=navyblue,
	pdftitle={Overleaf Example},
	pdfpagemode=FullScreen,
}
\usepackage{natbib}
\setcitestyle{authoryear}
\begin{document}

	\title{\bf Prediction intervals for quantile autoregression}
	\author{Silvia Novo{$^a$}\footnote{Corresponding author email address: \href{snovo@est-econ.uc3m.es}{snovo@est-econ.uc3m.es}} \hspace{3pt}
	C\'esar	S\'anchez-Sellero{$^b$} \\		
	{\normalsize $^a$ Department of Statistics, Universidad Carlos III de Madrid, Spain}\\
	{\normalsize $^b$ Department of Statistics, Mathematical Analysis and Optimization,}\\ {\normalsize Universidade de Santiago de Compostela, Spain}
	}
	
	\date{}
	\maketitle
	
	\begin{abstract} 
\noindent This paper introduces new methods for constructing prediction intervals using quantile-based techniques. The procedures are developed for both classical (homoscedastic) autoregressive models and modern quantile autoregressive models. They combine quantile estimation with multiplier bootstrap schemes to approximate the sampling variability of coefficient estimates, together with bootstrap replications of future observations. We consider both percentile-based and predictive-root-based constructions. Theoretical results establish the validity and pertinence of the proposed methods. Simulation experiments evaluate their finite-sample performance and show that the proposed methods yield improved coverage properties and computational efficiency relative to existing approaches in the literature. The empirical usefulness of the methods is illustrated through applications to U.S. unemployment rate data and retail gasoline prices.
	\end{abstract}
	
	\vspace{1cm}

	\section{Introduction}

	One of the main goals in time series analysis is to provide a prediction and a prediction interval for a future observation at time $(n+k)$, based on the available realization of the series up to time $n$. We will focus on autoregressive linear models, where new prediction intervals will be proposed using quantile estimation and bootstrap techniques.

    As a starting point, let us consider an observed series, $Y_1,\ldots,Y_n$, coming from a classical autoregressive model of order $p$, briefly AR($p$), of the form
\begin{equation} \label{eq:AR(p)}
	Y_t=\phi_0+\phi_1Y_{t-1}+\cdots+\phi_pY_{t-p}+a_t,\qquad t=1,\ldots,n,
\end{equation}
where $\phi_0,\phi_1,\ldots,\phi_p$ are unknown coefficients and innovations $a_t$ are independent random variables with a common distribution, $F_a$. In classical time series, the common distribution of the innovations is assumed to be centered normal with a certain variance $\sigma^2$.

Several methods to obtain prediction intervals for AR($p$) time series have been proposed in the literature, starting with the one given by \cite{box1976}, which strongly depends on the assumption that innovations follow a normal distribution. As a consequence, methods of this type are not suitable when this assumption is not satisfied. For this reason, several authors have proposed alternative bootstrap-based procedures. To understand the nature of these procedures, it is convenient to recall the prediction framework in autoregressive models. First, note that from the realization of the time series, $Y_1,\ldots,Y_n$, the coefficients in the model (\ref{eq:AR(p)}) can be estimated by least squares, as is commonly done in the literature, as follows:
$$\hat \phi=\arg \min_{\phi\in R^{p+1}}\sum_{t=p+1}^n \left(Y_t-\phi_0-\phi_1Y_{t-1}-\cdots-\phi_pY_{t-p}\right)^2,$$
where $\phi=(\phi_{0},\phi_1,\dots,\phi_p)^{\top}$.

Using estimated coefficients, predictions can be obtained by the recursion
$$\hat{Y}_{n+j}=\hat{\phi}_0+\hat{\phi}_1\hat{Y}_{n+j-1}+\dots+\hat{\phi}_p\hat{Y}_{n+j-p},\quad j=1,\ldots,k,$$
where $\hat{Y}_t=Y_t$ for $t=n-p+1,n-p+2,\dots,n$. Meanwhile, the true future values come from the same recursion with true coefficients and additional random innovations
$$Y_{n+j}=\phi_0+\phi_1 Y_{n+j-1}+\dots+\phi_p Y_{n+j-p}+a_{n+j},\quad j=1,\ldots,k,$$
where the innovations $a_{n+1},\ldots,a_{n+k}$ are independent random variables with the same common distribution, $F_a$, as in model (\ref{eq:AR(p)}).

Then, the prediction error, given by $Y_{n+j}-\hat{Y}_{n+j}$, has two sources of variability: one comes from the estimation of the coefficients and the other comes from the innovations corresponding to the future observation. To deal with these sources of variability, bootstrap methods generally adopt one of these two strategies (see \citealp{panpolitis2016}): percentile methods generate bootstrap replicas of future values, while other more recent methods generate bootstrap replicas of the so-called predictive root, $Y_{n+j}-\hat{Y}_{n+j}$.

In the group of percentile methods, \cite{stine1987} suggests to generate bootstrap values for the observed series and for future observations, using the estimated model and some bootstrap innovations, which are drawn from the empirical distribution of rescaled residuals. A drawback of \cite{stine1987}'s proposal is that it does not reproduce the randomness coming from the coefficients' estimation, which generally leads to undercoverage of prediction intervals. In order to include both the randomness of the future observation $Y_{n+k}$ conditioned on the observed series and the randomness of the coefficients' estimation, \cite{thombs1990} proposes to replicate the original series by means of the backward representation of the AR($p$) model. Noting that the future observation only depends on the last $p$ values of the observed series, the backward representation allows to fix these last $p$ observations while reproducing new values for the other elements of the series. The empirical distribution of the rescaled residuals, together with the coefficients estimated on the bootstrap series, is used to obtain bootstrap versions of the future observation. \cite{cao1997} proposed a simplification of \cite{thombs1990}'s method that omits the observed series generation. Their proposal saves computer time at the expense of a certain increase in coverage error. \cite{pascual2004} suggest to generate new values of the observed series with the common forward representation of the AR($p$) model. These new values are only used to obtain bootstrap replicates of the coefficients. Then, the last $p$ values of the series are fixed and used to generate bootstrap values of future observations. In this way, both the randomness from the coefficients' estimation and the randomness from the future observation are reproduced, without the need of a backward representation. \cite{pascual2004} made a theory based on M-estimation, which included estimators that minimize the least absolute deviation.

The predictive root, defined in \cite{politis2013}, where it was used to construct prediction intervals in regression, was adapted to autoregression models in \cite{panpolitis2016}, where several algorithms were given using forward and backward bootstrap and with different types of residual distributions and predictive roots.
	
In this work, two methods are proposed to obtain prediction intervals using quantile methods in classical autoregression models: one of the percentile style and the other one based on replicating a predictive root. The goal is to show that quantile estimation improves the performance of prediction intervals in autoregression models. First, recall that the coefficients of model (\ref{eq:AR(p)}) can be estimated by means of a quantile loss function as
\begin{equation} \label{eq:hatphi}
	\hat \phi=\arg \min_{\phi\in R^{p+1}}\sum_{t=p+1}^n \rho_\tau\left(Y_t-\phi_0-\phi_1Y_{t-1}-\cdots-\phi_pY_{t-p}\right),
\end{equation}
where $\phi=(\phi_{0},\phi_1,\dots,\phi_p)^{\top}$ and $\rho_\tau(u)=u(\tau-I(u<0))$ for $u\in R$ is the $\tau$-quantile loss function, with $\tau\in(0,1)$.

Quantile methods are receiving an increasing attention in the last decades (see \citealp{koenker2005,koenker2017}), due to their flexibility with respect to the error distribution (not necessarily normal) and possible heteroscedasticity. At the same time, quantile regression provides a more complete description of the response variable, because it allows to consider effects on central and non-central locations in the response variable. In time series, quantile methods lead to a more robust estimation of autoregression coefficients (see \citealp{herce1996}). In particular, least absolute deviation is the option primarily considered in the literature (see \citealp{breidt2001,pascual2004,li2008,wu2010}, among others), though other quantiles can be of interest depending on the expected shape of the error distribution.

To endow classical autoregression models with more generality, heteroscedastic components (see \citealp{engle1982,reeves2005,bose2009}), nonlinear (see \citealp{tong2011}) or even nonparametric models (see  \citealp{panpolitis2016}), and other types of autoregression dependencies have been studied in the literature. From a quantile perspective, quite general models can be given, as that introduced by \cite{koenker2006}. It is usually called as a QAR($p$) model, and can be stated as
\begin{equation}
	Y_t=\phi_0\left(U_t\right)+\phi_1\left(U_t\right)Y_{t-1}+\dots+\phi_p\left(U_t\right)Y_{t-p},
	\label{QAR_1}
\end{equation}
where $\phi_j:[0,1]\rightarrow R$ with $j=1,\dots,p$ are functions defined in the unit interval $[0,1]$ and $\{U_t\}_{t\in Z}$ is a sequence of independent random variables with uniform distribution on the interval $[0,1]$. Note that under this model the vector of coefficients  $\phi(\tau)=(\phi_{0}(\tau),\phi_1(\tau),\ldots,\phi_p(\tau))^{\top}$ depends on the quantile order $\tau$. Thus, this model is suited to heteroscedastic time series. It can also be observed that the coefficients are random, and that they are dependent.

The QAR($p$) model has attracted considerable attention in the literature, from both methodological (see \citealp{galvao2011,xiao2012,li2015,zhu2022,zhu2023}) and applied perspectives (see \citealp{ferreira2011,baur2012}, among others).
In particular, Section~9.1 of \cite{xiao2012} suggests a percentile-based procedure for constructing prediction intervals under the QAR($p$) model. 
However, this approach does not account for the additional uncertainty arising from the estimation of the model coefficients. To the best of our knowledge, this remains the only method available in the literature for constructing prediction intervals in the QAR($p$) framework. Motivated by this limitation, we propose  new methods for obtaining prediction intervals under the QAR($p$) model, including a percentile-based approach and a predictive-root-based approach, both accounting for the uncertainty associated with coefficient estimation.

The remainder of the paper is organized as follows. In Section~2, we introduce the proposed methods and discuss their theoretical properties. Section~3 presents a simulation study designed to analyze their finite-sample performance and to compare them with existing approaches in the literature. In Section~4, real data on the U.S. unemployment rate and the U.S. retail gasoline price are analyzed using the techniques proposed in this paper.
	
\section{The proposed methods}

\subsection{Prediction intervals for the classical autoregressive model}
    
In the framework of a classical autoregression model, as given in (\ref{eq:AR(p)}), but without the assumption of a normal distribution for innovations, a percentile method for obtaining prediction intervals is proposed, according to the following scheme.

\vspace*{5mm}
{\bf Algorithm AR-perc}

\begin{description}
	\item[\underline{Step 1}.]
	Fit the AR($p$) model given by (\ref{eq:AR(p)}) using a quantile loss function at a chosen order $\tau$. Most commonly, least absolute deviations will be used, corresponding to $\tau=0.5$. Let us denote by $\hat{\phi}$ the resulting estimations, as given by (\ref{eq:hatphi}). Obtain residuals as
    \begin{equation}
	   \hat{a}_t=Y_t-\hat{\phi}_0-\hat{\phi}_1Y_{t-1}-\dots-\hat{\phi}_pY_{t-p}, \quad t=p+1,p+2,\dots,n,
	   \label{residuos_qar}
    \end{equation}
    and compute their empirical distribution, that will be denoted by $\hat{F}_n$.
	\item[\underline{Step 2}.] Draw a sample of independent positive random variables $w_i$ with $i=1,\ldots,n$, such that $E(w_i)=1$ and $E(w_i^2)=2$. These random variables will also be independent of $\{Y_t\}_{t\in Z}$. For example, $w_i$ can be independent and exponentially distributed with mean $1$.
    
    Obtain a bootstrap estimate of $\phi$, by inserting the values $w_i$ as multipliers to the loss function, that is:
	\begin{eqnarray}
		\hat{\phi}^*=\arg\min_{\phi\in R^{p+1}}\sum_{t=p+1}^nw_t\rho_{\tau}(Y_t-\phi^{\top} Z_{t,p}),
		\label{estimacion_ponderada_proposta}
	\end{eqnarray}
    where $\phi=\left(\phi_0,\phi_1,\ldots,\phi_p\right)^T$ and $Z_{t,p}=\left(1,Y_{t-1},\ldots,Y_{t-p}\right)^T$.
    \item[\underline{Step 3}.] Draw bootstrap errors $a_{n+j}^*$ for $j=1,\dots,k$ independently from the empirical distribution of the residuals, $\hat{F}_n$, obtained in Step 1.

    Define $Y_t^*=Y_t$ for $t=n-p+1,n-p+2,\dots,n$ and compute future bootstrap observations following the expression:
	$$Y_{n+j}^*=\hat{\phi}_0^*+\hat{\phi}_1^*Y_{n+j-1}^*+\dots+\hat{\phi}_p^*Y_{n+j-p}^{*}+a_{n+j}^{*}\quad j=1,2,\dots,k.$$
\end{description}

To obtain the prediction intervals, Steps 2 and 3 are repeated $B$ times, where $B$ is a large enough number. One thousand is a common value (see \citealp{li2015,panpolitis2016}), necessary for a confidence level of 95\%. For higher levels, a larger $B$ would be required. The limits of the prediction intervals are obtained as the $\alpha/2$ and $(1-\alpha/2)$ quantiles of the empirical distribution of the bootstrap values, say $(Y_{n+k}^{*1},\ldots, Y_{n+k}^{*B})$.

Step 2 tries to mimic the randomness that comes from the estimation of the coefficients, while Step 3 replicates the randomness related to the evolution of the time series up to time $(n+k)$. Bootstrap multipliers were already used in the context of M-estimation in linear models by \cite{raozhao1992} and in quantile time series by \cite{li2015}. They have the advantage of better computational efficiency compared to bootstrap methods that generate replicas of the original sample of observations. Bootstrap multipliers also help to derive theoretical properties of the resulting prediction interval.

In Theorem \ref{th:AR-perc} the asymptotic validity of the Algorithm AR-perc is proven. First, the required assumption is established.

\begin{description}
    \item[(A. AR($p$))] The AR($p$) model given in (\ref{eq:AR(p)}) is assumed to be strictly stationary and causal, that is, $\phi(z)=1-\phi_1 z -\cdots-\phi_p z^p\neq 0$ for $|z|\leq 1$. Innovations are assumed to be independent random variables with a common distribution function, $F_a$, which has $\tau$-quantile equal to zero, finite variance $\sigma^2>0$, and a density that is bounded and bounded away from zero, that is, $m<f_a(x)<M$ for all $x$ such that $0<F_a(x)<1$, with $0<m<M<\infty$. 
\end{description}

Before the statement of Theorem \ref{th:AR-perc}, we show the strong consistency of the quantile estimator $\hat{\phi}$ and a Glivenko-Cantelli theorem of the empirical distribution $\hat{F}_n$. These results will be given in Lemma \ref{l:AR-perc}.

\begin{lemma} \label{l:AR-perc}{\it Let $\{Y_t\}_{t \in Z}$ be an AR($p$) process following the model given in (\ref{eq:AR(p)}) that satisfies the assumption (A. AR($p$)). Consider the estimators $\hat{\phi}$ and $\hat{F}_n$ obtained from the realization of the time series $(y_1,\ldots,y_n)$. Then, as $n$ goes to infinity, we have
	\begin{enumerate}
		\item[(a)] $\hat{\phi}$ converges almost surely to $\phi$.
        \item[(b)] $\sup_x\left|\hat F_n(x)-F_a(x)\right|$ converges to zero almost surely.
	\end{enumerate}
}
\end{lemma}

\begin{proof}
See the Appendix.
\end{proof}

\begin{theorem} \label{th:AR-perc} {\it Let $\{Y_t\}_{t \in Z}$ be an AR($p$) process following the model given in (\ref{eq:AR(p)}) that satisfies the assumption (A. AR($p$)). We have that, conditionally on the realization of the time series, $(y_1,\ldots,y_n)$:
	\begin{enumerate}
		\item[(a)] $\hat{\phi}^*$ converges in probability to $\phi$. 
		\item[(b)] $Y_{n+k}^*$ converges in distribution to $Y_{n+k}$.
	\end{enumerate}
}
\end{theorem}
\begin{proof}
 The proof of part (a) is similar to the proof of part (a) in Lemma \ref{l:AR-perc}. Here, we consider the bootstrap version of the objective function,
 $$R_n^*(v)=\frac{1}{n-p}\sum_{t=p+1}^n\omega_t\left[\rho_\tau\left(a_t-v^tZ_{t,p}\right)-\rho_\tau\left(a_t\right)\right],$$
where $v\in \mathbb{R}^{p+1}$ and $\omega_{p+1},\ldots,\omega_n$ are the bootstrap multipliers. Now, the minimizer of $R_n^*(v)$, say $\hat{v}^*$, will be $\hat{v}^*=\hat{\phi}^*-\phi$, so we have to prove that $\hat{v}^*$ converges to zero almost surely.

First, we decompose $R_n^*(v)$ into two summands,
$$R_n^*(v)=R_n(v)+\left(R_n^*(v)-R_n(v)\right).$$
 As in the proof of part (a) in Lemma \ref{l:AR-perc}, since $R_n^*(0)=0$ and $R_n^*(v)$ is a convex function (because the multipliers $\omega_t$ are non-negative), it suffices to show that for any $\Delta>0$,
\begin{equation} \label{eq:inf*}
    \liminf_{n\rightarrow\infty} \inf_{\|v\|=\Delta} R_n^*(v)>0\qquad\text{almost surely.}    
\end{equation}
 This fact was already proven in Lemma \ref{l:AR-perc} for $R_n(v)$ (stated in (\ref{eq:inf})), so we only have to show that $R_n^*(v)-R_n(v)$ converges uniformly to zero. To this end, first note that
$$R_n^*(v)-R_n(v)=\frac{1}{n-p}\sum_{t=p+1}^n\left(\omega_t-1\right)\left[\rho_\tau\left(a_t-v^tZ_{t,p}\right)-\rho_\tau\left(a_t\right)\right],$$
 so, conditionally to the realization of the time series $(y_1,\ldots,y_n)$, the addends in $R_n^*(v)-R_n(v)$ have conditional expectation equal to zero and are independent. Then, the mean will converge to zero almost surely because
 \begin{eqnarray*}
    \sum_{t=p+1}^\infty \frac{Var^*\left(\omega_t\left[\rho_\tau\left(a_t-v^tZ_{t,p}\right)-\rho_\tau\left(a_t\right)\right]\right)}{(t-p)^2}&=&\sum_{t=p+1}^\infty\frac{\left[\rho_\tau\left(a_t-v^tZ_{t,p}\right)-\rho_\tau\left(a_t\right)\right]^2}{(t-p)^2} \\
    &\leq& 4 \sum_{t=p+1}^\infty \frac{v^tZ_{t,p}Z_{t,p}^tv}{(t-p)^2}<\infty,     
 \end{eqnarray*}
 where $Var^*$ denotes the conditional variance. The first inequality comes from the fact that for any $a,b\in\mathbb{R}$, $\left|\rho_\tau(a-b)-\rho_\tau(a)\right|\leq 2 |b|$ . The last inequality, that is, the convergence of the series, is obtained from the stationarity of the AR($p$) process. Thus, it was proven that $R_n^*(v)-R_n(v)$ converges to zero almost surely for any $v\in\mathbb{R}^{p+1}$. This convergence is shown to be uniform if we observe that for any $v, v_0\in\mathbb{R}^{p+1}$
 $$\left|R_n^*(v)-R_n(v)-\left[R_n^*(v_0)-R_n(v_0)\right]\right| \leq \left|R_n^*(v)-R_n^*(v_0)\right|+\left|R_n(v)-R_n(v_0)\right|,$$
 and then recall the Lipschitz condition already obtained for $R_n(v)$ in the proof of Lemma \ref{l:AR-perc}, which can be derived for $R_n^*(v)$ with similar arguments.
 
   Part (b) can be proven following the same induction scheme given in the proof of Theorem 3.1 in \cite{thombs1990}, making use of part (a) of this theorem and part (b) of Lemma \ref{l:AR-perc}, which is used to show that $\hat{a}_t^*$ converges in distribution to $a_t$.
\end{proof}

An alternative way to obtain prediction intervals can be given based on bootstrapping the predictive root, $Y_{n+j}-\hat{Y}_{n+j}$. In addition, we use predictive residuals in Step 1 (see \cite{panpolitis2016}), instead of common residuals. All this is done in the next algorithm.

\vspace*{5mm}
{\bf Algorithm AR-proot}

\begin{description}
	\item[\underline{Step 1}.]
	Obtain predictive residuals as $Y_t-\hat{Y}_t$, where $\hat{Y}_t$ is the predicted value at time $t$ obtained by a fitted AR($p$) model. In this case, the model is again estimated by using a $\tau$-quantile loss function, but all terms involving the $Y_t$ observation are omitted. Thus, it is a leave-one-out procedure. The corresponding empirical distribution of predictive residuals is denoted by $\hat{F}_n^p$.
    \item[\underline{Step 2}.] Same as Step 2 in Algorithm AR-perc.
    \item[\underline{Step 3}.] From the bootstrap estimate of $\phi$, say $\hat{\phi}^*$, obtained in Step 2, bootstrap predicted values are computed following the recursion:
    $$\hat{Y}_{n+j}^*=\hat{\phi}_0^*+\hat{\phi}_1^*\hat{Y}_{n+j-1}^*+\dots+\hat{\phi}_p^*\hat{Y}_{n+j-p}^{*},\quad j=1,2,\dots,k,$$
    where $\hat{Y}_t^*=Y_t$ for $t=n-p+1,n-p+2,\dots,n$.
    
    Draw bootstrap errors $a_{n+j}^*$ for $j=1,\dots,k$ independently from the empirical distribution of the predictive residuals obtained in Step 1. Define $Y_t^*=Y_t$ for $t=n-p+1,n-p+2,\dots,n$ and compute future bootstrap observations following the expression:
	$$Y_{n+j}^*=\hat{\phi}_0+\hat{\phi}_1Y_{n+j-1}^*+\dots+\hat{\phi}_pY_{n+j-p}^{*}+a_{n+j}^{*},\quad j=1,2,\dots,k.$$
    Finally, the bootstrap replicate of the predictive root is computed as $Y_{n+j}^*-\hat{Y}_{n+j}^*$.
    \item[\underline{Step 4}.] Step 3 is run $B$ times and the empirical $\alpha$-quantile of the $B$ bootstrap predictive roots is denoted by $q(\alpha)$ for any $\alpha\in(0,1)$.
    \item[\underline{Step 5}.] Compute predicted future values by following the recursion
    $$\hat{Y}_{n+j}=\hat{\phi}_0+\hat{\phi}_1\hat{Y}_{n+j-1}+\dots+\hat{\phi}_p\hat{Y}_{n+j-p},\quad j=1,\ldots,k,$$
    where $\hat{Y}_t=Y_t$ for $t=n-p+1,n-p+2,\dots,n$.
    \item[\underline{Step 6}.] The equal-tailed prediction interval for $Y_{n+k}$ at the confidence level $(1-\alpha)$ is then constructed as
    $$\left[\hat{Y}_{n+k}+q(\alpha/2), \hat{Y}_{n+k}+q(1-\alpha/2)\right].$$
\end{description}

The Algorithm AR-proot can be shown to be pertinent in the sense of Definition 2.4 in \cite{panpolitis2016}. This is stated in Theorem \ref{th:AR-proot}. The concept of pertinence in this case applies to the following decomposition of the predictive root for lag $k=1$:
$$Y_{n+1}-\hat{Y}_{n+1}=A_n+a_{n+1},$$
where $A_n=\left(\phi_0-\hat{\phi}_0\right)+\left(\phi_1-\hat{\phi}_1\right)Y_n+\dots+\left(\phi_p-\hat{\phi}_p\right)Y_{n+1-p}$ represents the estimation error, while $a_{n+1}$ is the innovation coming from the new observation. A similar decomposition can be given for the bootstrap predictive root:
$$Y_{n+1}^*-\hat{Y}_{n+1}^*=A_n^*+a_{n+1}^*,$$
where $A_n^*=\left(\hat{\phi}_0-\hat{\phi}_0^*\right)+\left(\hat{\phi}_1-\hat{\phi}_1^*\right)Y_n+\dots+\left(\hat{\phi}_p-\hat{\phi}_p^*\right)Y_{n+1-p}$. 
An algorithm is then pertinent if $A_n^*$ approximates $A_n$, $a_{n+1}^*$ approximates $a_{n+1}$, and both components are independent. Note that pertinence is then a stronger condition than the validity of a prediction interval.

\begin{theorem}\label{th:AR-proot} {\it Let $\{Y_t\}_{t \in Z}$ be an AR($p$) process following the model given in (\ref{eq:AR(p)}) and satisfying the assumption (A. AR($p$)), we have that, conditionally on the realization of the time series, $(y_1,\ldots,y_n)$:
	\begin{enumerate}
		\item[(a)] $\left|P(\sqrt{n}A_n\leq a)-P^*(\sqrt{n}A_n^*\leq a)\right|$ converges to zero in probability, for all $a$.
        \item[(b)] $\sup_x\left|\hat F_n^p(x)-F_a(x)\right|$ converges to zero in probability.
		\item[(c)] $A_n^*$ and $a_{n+1}^*$ are independent.
	\end{enumerate}
}
\end{theorem}
\begin{proof}
 Part (a) can be derived from Corollary 2 in \cite{koenker2006} and Theorem 5 in \cite{li2015}, where the same limit distribution is obtained for $\sqrt{n}\left(\hat \phi-\phi\right)$ and $\sqrt{n}\left(\hat \phi^*-\hat\phi\right)$. Note that $A_n$ and $A_n^*$ are linear combinations of $(\hat\phi-\phi)$ and $(\hat\phi^*-\hat\phi)$, respectively. To prove part (b), we make use of \cite{panpolitis2016}, where it is shown that the difference between predictive residuals and common residuals is asymptotically negligible. This fact, together with our Lemma \ref{l:AR-perc}, proves part (b). 
 Part (c) is obvious from the construction of the Algorithm AR-proot.
\end{proof}

\subsection{Prediction intervals for the quantile autoregressive model}
 
For those situations that are not well adjusted by model (\ref{eq:AR(p)}), and where the more general model (\ref{QAR_1}) can be properly fitted, we propose two new methods to obtain prediction intervals: one of the percentile style and the other based on a predictive root. The following scheme provides a percentile method. 

\vspace*{5mm}
{\bf Algorithm QAR-perc}

\begin{description}
	\item[\underline{Step 1}.] Draw a sample of independent positive random variables $w_i$ with $i=1,\ldots,n$, such that $E(w_i)=1$ and $E(w_i^2)=2$. The random variables are also independent of $\{Y_t\}_{t\in Z}$.
	\item[\underline{Step 2}.] Draw independent random variables $U_{n+j}^*$ for $j=1,\ldots,k$, from the uniform distribution on the unit interval $[0,1]$. For each $U_{n+j}^*$, obtain a bootstrap version of coefficients using the multipliers $w_i$ (note that the same multipliers are used for all $U_{n+j}^*$'s):
	\begin{eqnarray}
		\hat{\phi}^*(U_{n+j}^*)=\arg\min_{\phi\in R^{p+1}}\sum_{t=p+1}^n w_t\rho_{U_{n+j}^*}\left(Y_t-\phi^{\top}Z_{t,p}^1\right).
	\end{eqnarray}
	\item[\underline{Step 3}.] Define $Y_t^*=Y_t$ for $t=n-p+1,n-p+2,\dots,n$ and compute future bootstrap observations by the recursion:
	$$Y_{n+j}^*=\hat{\phi}_0^*(U_{n+j}^*)+\hat{\phi}_1^*(U_{n+j}^*)Y_{n+j-1}^*+\dots+\hat{\phi}_p^*(U_{n+j}^*)Y_{n+j-p}^{*},\quad j=1,\dots,k.$$
	
\end{description}

Steps 1 to 3 are repeated $B$ times and the prediction interval is obtained similarly to the Algorithm AR-perc from the empirical distribution of the resulting bootstrap values.

The validity of the Algorithm QAR-perc is given in {Theorem \ref{th:QAR-perc}}.

\begin{theorem}\label{th:QAR-perc} {\it Let $\{Y_t\}_{t \in Z}$ be a QAR($p$) process following the model given in (\ref{QAR_1}) that satisfies assumptions of Theorem 5 in \cite{li2015}. We have that, conditionally on the realization of the time series, $(y_1,\ldots,y_n)$:
	\begin{enumerate}
		\item[(a)] $\sup_\tau\left|\hat{\phi}^*(\tau)-\phi(\tau)\right|$ converges in probability to zero.
        \item[(b)] $Y_{n+k}^*$ converges in distribution to $Y_{n+k}$.
	\end{enumerate}
}
\end{theorem}
\begin{proof}
 Part (a) can be proven from Theorem 2 in \cite{koenker2006}, where the process $\sqrt{n}\left(\hat{\phi}(\tau)-\phi(\tau)\right)$ is shown to converge in distribution to a Gaussian process and Theorem 5 in \cite{li2015} where the process $\sqrt{n}\left(\hat{\phi}^*(\tau)-\hat{\phi}(\tau)\right)$ is shown to converge in distribution to the same limit, conditionally on the realization of the time series.
 
 As in the Algorithm AR-perc, part (b) is obtained following the induction scheme given in the proof of Theorem 3.1 in \cite{thombs1990}. Then, we are going to provide the argument for $k=1$, while induction is applied for larger $k$. The prediction for $k=1$ can be decomposed as
 $$Y_{n+1}^*=Y_{n+1}^{0*}+R_{n+1}^*,$$
 where
$$Y_{n+1}^{0*}=\phi_0(U_{n+1}^*)+\phi_1(U_{n+1}^*)Y_n+\dots+\phi_p(U_{n+1}^*)Y_{n+1-p},$$
 and
 \begin{eqnarray*}
    R_{n+1}^*&=&\left(\hat{\phi}_0^*\left(U_{n+1}^*\right)-\phi_0\left(U_{n+1}^*\right)\right) \\
    &&+\left(\hat{\phi}_1^*\left(U_{n+1}^*\right)-\phi_1\left(U_{n+1}^*\right)\right)Y_{n}+\dots+\left(\hat{\phi}_p^*\left(U_{n+1}^*\right)-\phi_p\left(U_{n+1}^*\right)\right)Y_{n+1-p}.   
 \end{eqnarray*}
Note that, although the real random variable, $U_{n+1}$, representing the randomness for the new observation, may be different from its bootstrap replicate, $U_{n+1}^*$, both follow a uniform distribution. As a consequence, $Y_{n+1}^{0*}$ follows the same distribution as $Y_{n+1}$.

Part (a) can be used to prove that $R_{n+1}^*$ converges in probability to zero. Applying Slutsky's theorem, $Y_{n+1}^*$ is shown to converge in distribution to $Y_{n+1}$.
\end{proof}

Alternatively to the previous percentile interval, we propose a method based on a prediction for a certain quantile order, $\tau_0$, and its corresponding predictive root. Let $\hat{Y}_t=Y_t$ for $t=n-p+1,n-p+2,\dots,n$. The prediction is obtained by the recursion
\begin{equation} \label{eq:pred_QAR}
\hat{Y}_{n+j}=\hat{\phi}_0\left(\tau_0\right)+\hat{\phi}_1\left(\tau_0\right)\hat{Y}_{n+j-1}+\dots+\hat{\phi}_p\left(\tau_0\right)\hat{Y}_{n+j-p},\quad j=1,\ldots,k,
\end{equation}
where $\hat{\phi}(\tau_0)=\left(\hat{\phi}_0(\tau_0),\hat{\phi}_1(\tau_0),\ldots,\hat{\phi}_p(\tau_0)\right)$ is obtained from (\ref{eq:hatphi}). Then, the $k$-step predictive root is defined as $Y_{n+k}-\hat{Y}_{n+k}$. Now, the algorithm can be described in the following steps.

\vspace*{5mm}
{\bf Algorithm QAR-proot}

\begin{description}
    \item[\underline{Step 1}.] Fit the QAR($p$) model given by (\ref{QAR_1}) at the chosen quantile order $\tau_0$ and obtain the prediction $\hat{Y}_{n+k}$ as shown in (\ref{eq:pred_QAR}).
    \item[\underline{Step 2}.] A sample of multipliers $w_i$ is drawn as described in Step 1 of Algorithm QAR-perc, and used to obtain a bootstrap version of coefficients at $\tau_0$:
	\begin{eqnarray}
		\hat{\phi}^*(\tau_0)=\arg\min_{\phi\in R^{p+1}}\sum_{t=p+1}^n w_t\rho_{\tau_0}\left(Y_t-\phi^{\top}Z_{t,p}^1\right).
	\end{eqnarray}
    \item[\underline{Step 3}.] Bootstrap predictive roots are computed as follows.
    \begin{enumerate}
        \item Define $\hat{Y}_t^*=Y_t$ for $t=n-p+1,n-p+2,\dots,n$ and compute the future bootstrap predicted values by the recursion:
	$$\hat{Y}_{n+j}^*=\hat{\phi}_0^*(\tau_0)+\hat{\phi}_1^*(\tau_0)\hat{Y}_{n+j-1}^*+\dots+\hat{\phi}_p^*(\tau_0)\hat{Y}_{n+j-p}^{*},\quad j=1,\dots,k.$$
        \item Draw independent random variables $U_{n+j}^*$ for $j=1,\ldots,k$, from the uniform distribution on the unit interval $[0,1]$. Let $Y_t^*=Y_t$ for $t=n-p+1,n-p+2,\dots,n$ and compute the future bootstrap values by the recursion:
	$$Y_{n+j}^*=\hat{\phi}_0(U_{n+j}^*)+\hat{\phi}_1(U_{n+j}^*)Y_{n+j-1}^*+\dots+\hat{\phi}_p(U_{n+j}^*) Y_{n+j-p}^{*},\quad j=1,\dots,k,$$
        where $\hat{\phi}(U_{n+j}^*)=\left(\hat{\phi}_0(U_{n+j}^*),\hat{\phi}_1(U_{n+j}^*),\ldots,\hat{\phi}_p(U_{n+j}^*)\right)$ is obtained from (\ref{eq:hatphi}) with $\tau=U_{n+j}^*$.
        \item Compute the bootstrap predictive root as $Y_{n+k}^*-\hat{Y}_{n+k}^*$.
    \end{enumerate}
    \item[\underline{Step 4}.] Steps 2 and 3 are run $B$ times and the empirical $\alpha$-quantile of the $B$ bootstrap predictive roots is denoted by $q(\alpha)$ for any $\alpha\in(0,1)$.
    \item[\underline{Step 5}.] The equal-tailed prediction interval for $Y_{n+k}$ at the confidence level $(1-\alpha)$ is then constructed as
    $$\left[\hat{Y}_{n+k}+q(\alpha/2), \hat{Y}_{n+k}+q(1-\alpha/2)\right].$$
	
\end{description}

Note that, differently from Step 2 in Algorithm QAR-perc, here the bootstrap version of the coefficients is only computed for the quantile order $\tau_0$. This fact provides some advantages from the theoretical point of view and could improve its statistical performance.

Regarding the pertinence of Algorithm QAR-proot, let us consider the decomposition of the predictive root for lag $k=1$:
$$Y_{n+1}-\hat{Y}_{n+1}=A_n+a_{n+1},$$
where
$$A_n=\left(\phi_0(\tau_0)-\hat{\phi}_0(\tau_0)\right)+\left(\phi_1(\tau_0)-\hat{\phi}_1(\tau_0)\right)Y_n+\dots+\left(\phi_p(\tau_0)-\hat{\phi}_p(\tau_0)\right)Y_{n+1-p},$$
represents the estimation error while $$a_{n+1}=\left(\phi_0(U_{n+1})-\phi_0(\tau_0)\right)+\left(\phi_1(U_{n+1})-\phi_1(\tau_0)\right)Y_n+\dots+\left(\phi_p(U_{n+1})-\phi_p(\tau_0)\right)Y_{n+1-p},$$
shows the randomness coming from the new observation. Similarly, the bootstrap predictive root can be written:
$$Y_{n+1}^*-\hat{Y}_{n+1}^*=A_n^*+a_{n+1}^*,$$
where
$$A_n^*=\left(\hat{\phi}_0(\tau_0)-\hat{\phi}_0^*(\tau_0)\right)+\left(\hat{\phi}_1(\tau_0)-\hat{\phi}_1^*(\tau_0)\right)Y_n+\dots+\left(\hat{\phi}_p(\tau_0)-\hat{\phi}_p^*(\tau_0)\right)Y_{n+1-p},$$
and
$$a_{n+1}^*=\left(\hat{\phi}_0(U_{n+1}^*)-\hat{\phi}_0(\tau_0)\right)+\left(\hat{\phi}_1(U_{n+1}^*)-\hat{\phi}_1(\tau_0)\right)Y_n+\dots+\left(\hat{\phi}_p(U_{n+1}^*)-\hat{\phi}_p(\tau_0)\right)Y_{n+1-p}.$$

Theorem \ref{th:QAR-proot} asserts the pertinence of Algorithm QAR-proot, by showing that $A_n^*$ approximates $A_n$, $a_{n+1}^*$ approximates $a_{n+1}$, and both components of the bootstrap predictive root are independent.

\begin{theorem} \label{th:QAR-proot} {\it Let $\{Y_t\}_{t \in Z}$ be a QAR($p$) process following the model given in (\ref{QAR_1}) that satisfies assumptions of Theorem 5 in \cite{li2015}. We have that, conditionally on the realization of the time series, $(y_1,\ldots,y_n)$:
	\begin{enumerate}
		\item[(a)] $\left|P(\sqrt{n}A_n\leq a) -P^*(\sqrt{n}A_n^*\leq a)\right|$ converges to zero in probability, for all $a$.
        \item[(b)] $\sup_a\left|P^*(a_{n+1}^*\leq a)-P(a_{n+1}\leq a)\right|$ converges to zero in probability.
		\item[(c)] $A_n^*$ and $a_{n+1}^*$ are independent.
	\end{enumerate}
}
\end{theorem}
\begin{proof}
Part (a) can be obtained from Theorem 2 in \cite{koenker2006} and Theorem 5 in \cite{li2015}, noting that $A_n$ and $A_n^*$ are linear combinations of $(\hat\phi(\tau_0)-\phi(\tau_0))$ and $(\hat\phi^*(\tau_0)-\hat\phi(\tau_0))$, respectively.

For part (b) we consider the following decomposition
$$a_{n+1}^*=a_{n+1}^{0*}+S_{n+1}^{*}(U_{n+1}^*)+S_{n+1}^{*}(\tau_0),$$
where
$$a_{n+1}^{0*}=\left(\phi_0(U_{n+1}^*)-\phi_0(\tau_0)\right)+\left(\phi_1(U_{n+1}^*)-\phi_1(\tau_0)\right)Y_n+\dots+\left(\phi_p(U_{n+1}^*)-\phi_p(\tau_0)\right)Y_{n+1-p},$$
and
$$S_{n+1}^{*}(\tau)=\left(\hat{\phi}_0(\tau)-\phi_0(\tau)\right)+\left(\hat{\phi}_1(\tau)-\phi_1(\tau)\right)Y_n+\dots+\left(\hat{\phi}_p(\tau)-\phi_p(\tau)\right)Y_{n+1-p},$$
for any $\tau\in(0,1)$. The quantity $a_{n+1}^{0*}$ follows the same distribution as $a_{n+1}$, while $S_{n+1}^*(\tau)$ converges to zero uniformly with respect to $\tau$, so Slutsky's theorem concludes the proof of part (b).

Part (c) is easily derived from the construction of the Algorithm QAR-proot.
\end{proof}

\vspace*{3mm}

\section{Simulations}
This section presents a simulation study to assess the performance of the methods proposed for the AR($p$) (\ref{eq:AR(p)}) and QAR($p$) (\ref{QAR_1}) models. Although the four procedures are asymptotically valid under the corresponding models, the conditional coverage of the methods with finite samples may be affected by the configuration of the model, the parameter estimation and the innovations' distribution. Consequently, we designed a complete simulation study to analyze the role of these settings 
in the conditional coverage produced by the methods.
\subsection{The design} 
The study focuses on the following simulated models:
\begin{description}
\item[Model 1:] $Y_t=\phi_1Y_{t-1}+a_t$. We varied the value of the autoregressive coefficient
as follows: $\phi_1=\ell/10$ with $\ell=1,\dots,9$. 
\item[Model 2:] $Y_t=0.75Y_{t-1}+\sum_{v=2}^p(-1)^{v-1}0.50Y_{t-v}+a_t$. We take $p=2,6$. The case $p=2$ gives rise to a model that was used in \cite{thombs1990} and \cite{cao1997}  to assess the performance of their methods.

\item[Model 3:] $Y_t=\phi_0(U_t)+\phi_1(U_t)Y_{t-1}$ where $\phi_0(U_t)=F_a^{-1}(U_t)$ and $\phi_1(U_t)=\min\{\gamma_0+\gamma_1F_{a}(U_t),1\}$. We took $\gamma_0=0.25$ and $\gamma_1=0.85$. This type of model was analyzed by \cite{koenker2006}. 

 \item[Model 4:] $Y_t=\phi_0(U_t)+\phi_1(U_t)Y_{t-1}+\phi_2(U_t)Y_{t-2}$ where $\phi_0(U_t)=F_a^{-1}(U_t)$, $\phi_1(U_t)=0.3$, and $\phi_2(U_t)=0.7F_a(U_t)$. 
\end{description}
Models 1 and 2 are AR($p$) processes, while Models 3 and 4 correspond to QAR($p$) processes.
Three distributions were considered for the innovations $a_t$ in Models 1 and 2 and for the distribution function $F_a$ in Models 3 and 4: 
standard normal (N$(0,1)$), Student's T with 3 degrees of freedom ($T_3$), and chi-squared with 5 degrees of freedom ($\chi^2_5$). 
The sample sizes were $n=25, 50, 100, 200, 300, 1000$, and the prediction horizons were $k=1, 2, 3,4$. The nominal confidence levels $\beta=1-\alpha$ were established at $\beta=0.90$ and $\beta=0.95$. 

A total of $S=500$ time series were generated from Models 1--4 with the true coefficients, error distributions and sample sizes. In all cases, an initial burn-in period of 300 observations was discarded in order to ensure that the simulated series had reached their stationary regime.
For each simulated series, every method produces a $100\beta\%$ prediction interval at each horizon 1, 2, 3 and 4. 
Let us denote one such interval by $(L_s,U_s)$ for $s=1,\ldots,S$. In order to evaluate the performance of each method, $F=1000$ future values are drawn from the true model, simulated series, and prediction horizon. Let us denote them by $Y_{n+k}^{s,f}$ for $s=1,\ldots,S$, $f=1,\ldots,F$ and $k=1,2,3,4$.
Then, the coverage conditioned to the $s$-th observed series for a particular method is given by $\beta_s=P(L_s<Y_{n+k}^{s,f}<U_s\mid Y_{1}^s,\dots,Y_{n}^s)$ and can be estimated by
\begin{equation}\hat{\beta}_{s}=\frac{1}{F}\sum_{f=1}^FI\left(L_s<Y_{n+k}^{s,f}<U_s\right).\label{eq:cov}\end{equation}
From the sample of conditional coverages $\hat{\beta}_s$ for $s=1,\dots, S$, we obtained more stable statistics, such as the average, standard error, and mean squared error of the coverage, as follows:
\begin{equation*}\overline{\hat{\beta}}=\frac{1}{S}\sum_{s=1}^S \hat{\beta}_s, \quad {SE}\left(\overline{\hat{\beta}}\right)=\frac{1}{\sqrt{S}}\sqrt{\frac{\sum_{s=1}^S \left(\hat{\beta}_s-\overline{\hat{\beta}}\right)^2}{S-1}}, \quad
	{MSE}=\frac{1}{S}\sum_{s=1}^S\left(\hat{\beta}_s-\beta\right)^2.
\end{equation*} 
 Clearly, a good method should produce average coverages close to the nominal confidence level $\beta$ with low SE and MSE. 
Moreover, the distribution of the conditional coverage should be symmetric around the nominal level. This means, first, that the median should also align with the nominal value, i.e. $\gamma=P(\beta_s\geq\beta)=0.5$, leading us to define a fourth summary statistic, $\hat{\gamma}$, which we expect to be close to 0.5: \begin{equation*}\hat{\gamma}=\frac{1}{S}\sum_{s=1}^SI\left(\hat{\beta}_s\geq \beta\right).\end{equation*}
Second,  from the  $100\alpha_s\%=100(1-\beta_s)\%$ that fall outside the produced interval, one half should fall below and the other half above it.
This allows us to consider the following statistics to complete our study,
\begin{equation*}\hat{A}_s=\frac{1}{F}\sum_{f=1}^FI\left(Y_{n+k}^{s,f}>U_s\right),
\quad \hat{B}_s=\frac{1}{F}\sum_{f=1}^FI\left(Y_{n+k}^{s,f}<L_s\right),\end{equation*}
and compute their sample averages:
\begin{equation*}\overline{\hat{A}}=\frac{1}{S}\sum_{s=1}^S \hat{A}_s, \quad \overline{\hat{B}}=\frac{1}{S}\sum_{s=1}^S \hat{B}_s.
\end{equation*}

Finally, we can also calculate the length of intervals and obtain their sample average and standard error: \begin{equation*}
\textrm{Len}_{s}=U_s-L_s, \quad \overline{\textrm{Len}}=\frac{1}{S}\sum_{s=1}^S\textrm{Len}_{s}, \quad   SE\left(\overline{\textrm{Len}}\right)=\frac{1}{S}\sqrt{\frac{\sum_{s=1}^S \left(\textrm{Len}_s-\overline{\textrm{Len}}\right)^2}{S-1}}. \end{equation*} A short length will clearly be preferable.

The performance of the proposed procedures for AR($p$) models (AR-perc and AR-proot) and QAR($p$) models (QAR-perc and QAR-proot) is assessed by comparing them with several competing approaches from the existing literature. Specifically, we consider the classical approach of \cite{box1976} (BJ) for the AR($p$) model and a range of bootstrap-based methods, which can be grouped as follows:
\begin{description}
\item[AR($p$)-based competitor methods.] Bootstrap procedures constructed under the AR($p$) framework, using either the percentile or the predictive root approach:
\begin{itemize}
\item Percentile methods:
\begin{itemize}
\item the backward bootstrap proposed by \cite{thombs1990} (TS),
\item the bootstrap of \cite{cao1997} (CB),
\item the forward bootstrap of \cite{pascual2004}, implemented using least squares (PRR) and least absolute deviation estimation (PRR-LAD).
\end{itemize}
\item Predictive root methods:
\begin{itemize}
\item the forward bootstrap with predictive residuals proposed by \cite{panpolitis2016} (PP).
\end{itemize}
\end{itemize}
\item[QAR($p$)-based competitor methods.] To the best of our knowledge, the only prediction interval method available in the literature for the QAR($p$) model is the percentile bootstrap suggested by \cite{xiao2012}.
\end{description}
In the simulation study, we set the number of bootstrap replications to $B=1000$ for all AR($p$)-based methods and to $B=5000$ for all QAR($p$)-based methods, which was found to be sufficient in practice.

Additionally, for each simulated time series 
$s$, with 
$s=1,\dots, S$, we complement the evaluation of the methods by constructing empirical prediction intervals through forward replication of the data-generating process (ORACLE). Specifically, each series is extended into the future by repeatedly simulating the true model, which allows us to obtain prediction intervals with exact nominal confidence level. Although such oracle intervals are infeasible in practical applications, they provide a natural benchmark for assessing the expected length of prediction intervals under each scenario.

For ease of interpretation, all reported values of average coverage, its standard error, mean squared error of coverage, as well as $\overline{\hat{A}}$ and $\overline{\hat{B}}$, are expressed in percentage terms in the tables and figures.

\subsection{The results} 

\subsubsection{Model 1}
The simulation study includes Model 1 to examine the impact of the autoregressive coefficient in an AR(1) process on the coverage performance of each method. 

Since we consider values of $\phi_1$ within the interval $(0,1)$, the generated autoregressive process is stationary for each value of $\phi_1$. However, as $\phi_1$ increases, the unconditional variance of $Y_t$, given by $\text{Var}(Y_t) = \sigma^2/(1 - \phi_1^2)$, also increases, bringing the process closer to nonstationarity.

\begin{figure}[H]
\centering
\includegraphics[width=17.3cm, keepaspectratio=true]{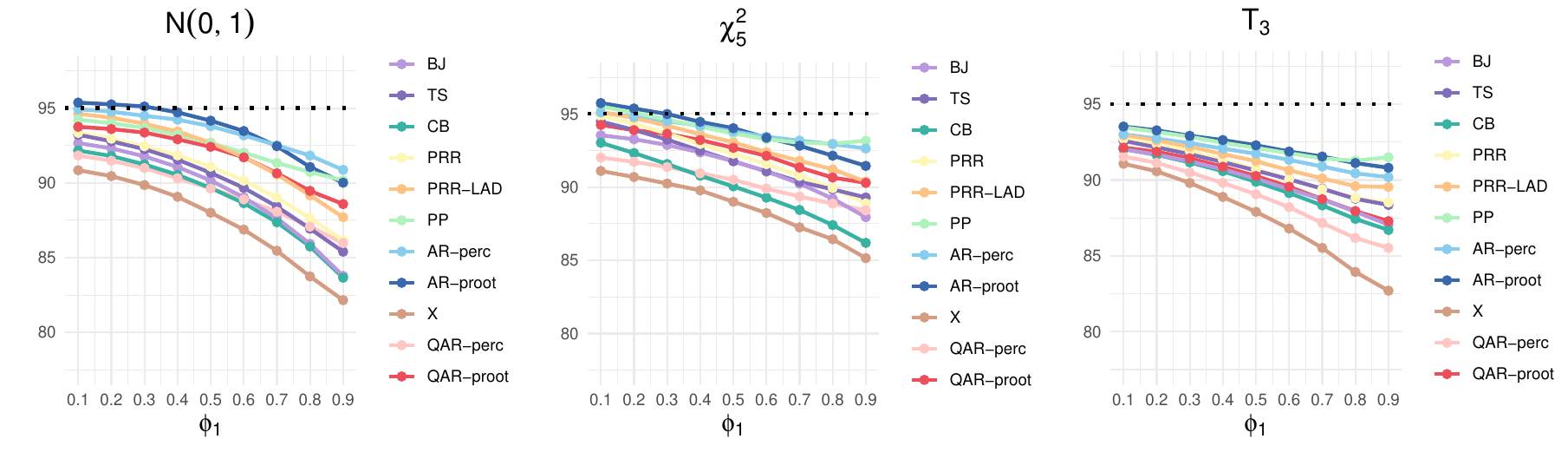}
\includegraphics[width=17.3cm, keepaspectratio=true]{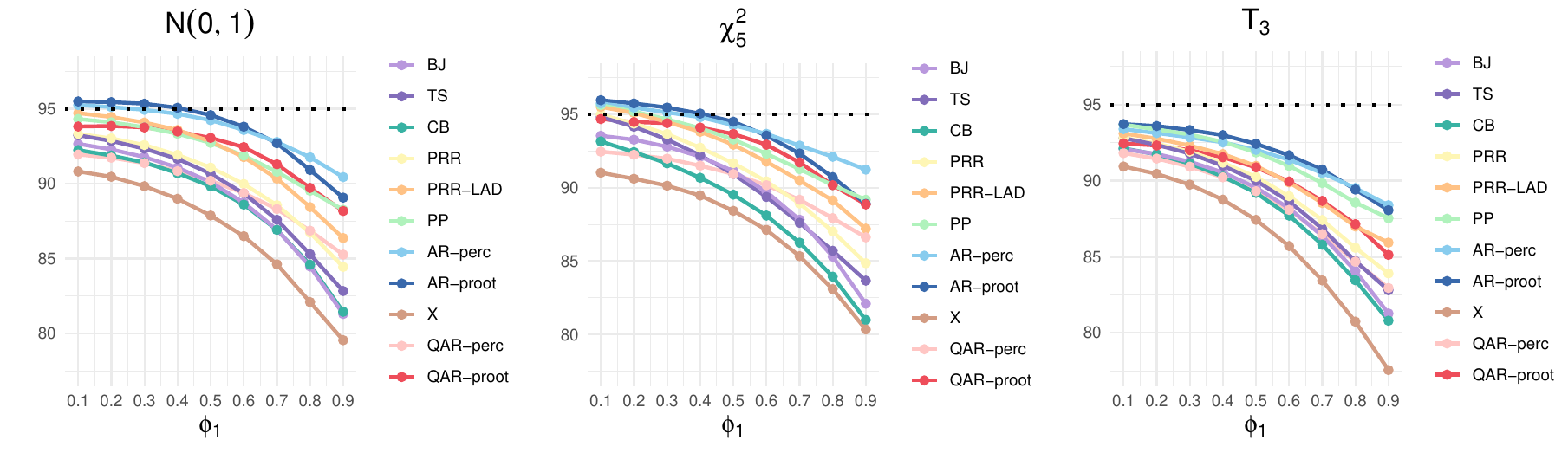}
\caption{Average coverage $\overline{\hat{\beta}}$ for each method as a function of $\phi_1$ in Model~1 with $n=25$ and $S=500$. Results are shown for  $N(0,1)$, $\chi^2_5$, and $T_3$ innovations, with horizons $k=2$ (top row) and $k=4$ (bottom row). The nominal confidence level is $\beta=0.95$ (dotted line).
} 
	\label{fig:ar1}
\end{figure}

Figure \ref{fig:ar1} shows the average coverage of the methods, $\overline{\hat{\beta}}$, across different values of \( \phi_1 \) in Model 1 for \( n=25 \) and horizon $k=2$ (above) and $k=4$ (below). N$(0,1)$, $T_3$ and $\chi^2_5$ innovations are considered. 
The nominal confidence level was $\beta=0.95$. We observe that an increase in the autoregressive coefficient leads to a deterioration in the coverage produced by all procedures, more evident when $k=4$. The AR-proot and AR-perc methods (represented by the blue and light blue lines in Figure~\ref{fig:ar1}, respectively) demonstrate the best overall performance, yielding similar average coverage rates. For cases where $\phi_1 > 0.7$, AR-perc achieves superior results for $N(0,1)$ and $\chi_5^2$ innovations, while for $T_3$ innovations AR-proot slightly outperforms. The results also indicate that the QAR-perc and, in particular, the QAR-proot methods, despite their greater complexity,  produce average coverages comparable to those obtained by methods developed for AR($p$) models.

\begin{table}[H]
\centering
	\scalebox{0.8}{
		\begin{tabular}{|c|r|rrrrr|rrrrr|}
			\cline{3-12}
			\multicolumn{2}{c|}{}& \multicolumn{5}{c}{N(0,1)} &\multicolumn{5}{|c|}{$\chi^2_5$}\\
			\hline
        n	& Method & $\overline{\hat{\beta}}$ (SE) & {MSE}& $\overline{\hat{B}}$/$\overline{\hat{A}}$ & $\overline{\textrm{Len}}$(SE) & $\hat{\gamma}$ & $\overline{\hat{\beta}}$ (SE) & {MSE}& $\overline{\hat{B}}$/$\overline{\hat{A}}$ & $\overline{\textrm{Len}}$(SE) & $\hat{\gamma}$ \\ 
        	\hline
            &&&&&&&&&&&\\
&ORACLE & 95.00 (0.00) & 0.00 & 2.50/2.50 & 3.92 (0.12) & 0.50 & 95.00 (0.00) & 0.00 & 2.50/2.50 & 11.95 (0.47) & 0.50 \\
&BJ & 91.34 (0.24) & 0.43 & 4.57/4.09 & 3.72 (0.03) & 0.27 & 92.79 (0.24) & 0.32 & 0.53/6.68 & 11.76 (0.11) & 0.38 \\
&TS & 91.93 (0.25) & 0.41 & 4.25/3.82 & 3.95 (0.03) & 0.34 & 93.35 (0.28) & 0.42 & 2.55/4.10 & 12.66 (0.15) & 0.50 \\
&CB & 90.86 (0.29) & 0.60 & 4.89/4.24 & 3.87 (0.03) & 0.29 & 91.32 (0.33) & 0.67 & 4.51/4.17 & 12.21 (0.14) & 0.37 \\
&PRR & 91.85 (0.26) & 0.44 & 4.30/3.85 & 3.98 (0.03) & 0.36 & 93.47 (0.26) & 0.37 & 2.37/4.15 & 12.72 (0.15) & 0.50 \\
&PRR-LAD & 92.80 (0.25) & 0.36 & 3.87/3.33 & 4.20 (0.03) & 0.41 & 93.80 (0.25) & 0.32 & 2.44/3.76 & 13.08 (0.14) & 0.52 \\
25&PP & 93.78 (0.22) & 0.26 & 3.24/2.98 & 4.29 (0.03) & 0.47 & 94.90 (0.26) & 0.33 & 1.75/3.36 & 13.78 (0.16) & 0.65 \\
&AR-perc& 93.18 (0.23) & 0.30 & 3.61/3.21 & 4.18 (0.04) & 0.42 & 94.09 (0.25) & 0.31 & 2.13/3.78 & 13.18 (0.16) & 0.55 \\
&AR-proot & 93.58 (0.25) & 0.33 & 3.55/2.87 & 4.40 (0.04) & 0.50 & 94.50 (0.27) & 0.37 & 1.81/3.69 & 13.90 (0.17) & 0.60 \\
&X & 85.38 (0.45) & 1.93 & 7.60/7.02 & 3.48 (0.05) & 0.14 & 86.33 (0.48) & 1.88 & 6.73/6.94 & 10.91 (0.18) & 0.17 \\
&QAR-perc & 86.94 (0.39) & 1.39 & 6.90/6.16 & 3.58 (0.05) & 0.17 & 88.35 (0.37) & 1.14 & 5.60/6.06 & 11.40 (0.19) & 0.22 \\
&QAR-proot & 90.35 (0.34) & 0.78 & 5.14/4.51 & 3.98 (0.05) & 0.32 & 92.46 (0.31) & 0.54 & 2.21/5.33 & 12.69 (0.21) & 0.44 \\
&&&&&&&&&&&\\
\hline
&&&&&&&&&&&\\
& ORACLE & 95.00 (0.00) & 0.00 & 2.50/2.50 & 3.92 (0.12) & 0.50 & 95.00 (0.00) & 0.00 & 2.50/2.50 & 11.97 (0.50) & 0.50 \\
& BJ& 93.26 (0.14) & 0.13 & 3.46/3.29 & 3.81 (0.02) & 0.31 & 94.26 (0.16) & 0.14 & 0.22/5.53 & 12.16 (0.08) & 0.46 \\
& TS & 93.44 (0.15) & 0.14 & 3.22/3.34 & 3.94 (0.02) & 0.35 & 94.24 (0.21) & 0.22 & 2.50/3.25 & 12.33 (0.10) & 0.52 \\
& CB & 92.21 (0.19) & 0.25 & 3.84/3.95 & 3.82 (0.03) & 0.27 & 92.57 (0.24) & 0.36 & 3.89/3.55 & 12.00 (0.12) & 0.34 \\
& PRR & 93.49 (0.16) & 0.14 & 3.19/3.32 & 3.96 (0.02) & 0.36 & 94.42 (0.18) & 0.17 & 2.35/3.24 & 12.50 (0.11) & 0.53 \\
& PRR-LAD & 93.97 (0.15) & 0.12 & 2.91/3.11 & 4.05 (0.02) & 0.43 & 94.50 (0.18) & 0.16 & 2.45/3.04 & 12.71 (0.11) & 0.54 \\
50& PP& 94.28 (0.15) & 0.11 & 2.81/2.90 & 4.10 (0.02) & 0.48 & 95.54 (0.19) & 0.19 & 1.76/2.70 & 13.14 (0.11) & 0.65 \\
& AR-perc & 93.88 (0.15) & 0.13 & 2.97/3.16 & 4.03 (0.02) & 0.43 & 94.66 (0.18) & 0.16 & 2.21/3.13 & 12.70 (0.12) & 0.56 \\
& AR-proot & 94.26 (0.15) & 0.12 & 2.78/2.96 & 4.13 (0.02) & 0.48 & 95.16 (0.18) & 0.16 & 1.73/3.11 & 13.03 (0.12) & 0.61 \\
 & X & 90.86 (0.27) & 0.53 & 4.69/4.45 & 3.77 (0.03) & 0.25 & 91.17 (0.30) & 0.58 & 4.50/4.33 & 11.90 (0.15) & 0.30 \\
&QAR-perc & 91.70  (0.24) & 0.40 & 3.92/4.38 & 3.87 (0.03) & 0.28 & 91.95 (0.26) & 0.42 & 3.79/4.26 & 11.94 (0.15) & 0.31 \\
& QAR-proot & 93.01 (0.21) & 0.26 & 3.34/3.64 & 4.02 (0.03) & 0.38 & 94.19 (0.21) & 0.23 & 1.75/4.06 & 12.51 (0.15) & 0.53 \\
&&&&&&&&&&&\\
\hline
\end{tabular}}
\caption{Simulation results for Model~1 with $\phi_1=0.6$ and horizon $k=1$, for $n=25$ and $n=50$ with $S=500$. The table reports average coverage $\overline{\hat{\beta}}$ (with its standard error),  mean squared error of the coverage (MSE), tail probabilities $\overline{\hat{B}}$/$\overline{\hat{A}}$, average interval length (with its standard error) and symmetry statistic $\hat{\gamma}$ for $N(0,1)$ and $\chi^2_5$ innovations. The nominal confidence level is $\beta=0.95$.
}
\label{tab:ar1}
\end{table}

To further analyze the prediction intervals produced by the methods, we selected $\phi_1=0.6$ and summarized the metrics in Table \ref{tab:ar1} and Table \ref{tab:ar12}. For clarity, Table \ref{tab:ar1} only includes the prediction horizon $k=1$,  sample sizes $n=25, 50$, N$(0,1)$ and $\chi_5^2$ innovations and nominal confidence level $\beta=0.95$. Table \ref{tab:ar12} presents the same settings for $k=3$. 
Tables~\ref{tab:ar1} and~\ref{tab:ar12} show that all methods improve their average coverage and MSE when increasing the sample size from $n = 25$ to $n = 50$, which suggests that they are consistent.
As shown in Table~\ref{tab:ar1}, for $k=1$, PP attains average coverages closest to the nominal level, but AR-proot displays a comparable performance in terms of average coverage, with differences that are not statistically significant at the 5\% level. For $k = 3$, Table~\ref{tab:ar12} indicates that  AR-perc and AR-proot are the best performing methods, achieving the average coverages closest to the nominal level across all scenarios, while also yielding the lowest MSE values. Both methods perform similarly for $k = 3$. Moreover, the differences in coverage between these procedures and the next-best method (PP) are statistically significant. In contrast, the X method performs the worst for both $k=1$ and $k=3$, exhibiting poor coverage for small sample sizes.

\begin{table}[H]
\centering
	\scalebox{0.8}{
		\begin{tabular}{|c|r|rrrrr|rrrrr|}
			\cline{3-12}
			\multicolumn{2}{c|}{}& \multicolumn{5}{c}{N(0,1)} &\multicolumn{5}{|c|}{$\chi^2_5$}\\
			\hline
        n	& Method & $\overline{\hat{\beta}}$ (SE) & {MSE}& $\overline{\hat{B}}$/$\overline{\hat{A}}$ & $\overline{\textrm{Len}}$(SE) & $\hat{\gamma}$ & $\overline{\hat{\beta}}$ (SE) & {MSE}& $\overline{\hat{B}}$/$\overline{\hat{A}}$ & $\overline{\textrm{Len}}$(SE) & $\hat{\gamma}$ \\ 
        	\hline
            &&&&&&&&&&&\\
	&   ORACLE & 95.00 (0.00) & 0.00 & 2.50/2.50 & 4.78 (0.14) & 0.50 & 95.00 (0.00) & 0.00 & 2.50/2.50 & 14.90 (0.54) & 0.50 \\ 
 	&    BJ & 89.00 (0.32) & 0.88 & 5.80/5.20 & 4.35 (0.04) & 0.22 & 90.15 (0.33) & 0.78 & 2.31/7.55 & 13.73 (0.15) & 0.30 \\ 
 	&    TS & 89.66 (0.32) & 0.79 & 5.38/4.96 & 4.51 (0.04) & 0.25 & 89.93 (0.34) & 0.84 & 4.75/5.32 & 14.40 (0.16) & 0.30 \\ 
  	&   CB & 88.66 (0.35) & 1.00 & 5.88/5.46 & 4.39 (0.04) & 0.22 & 88.40 (0.36) & 1.10 & 6.03/5.57 & 13.81 (0.15) & 0.21 \\ 
 	&    PRR & 90.14 (0.32) & 0.75 & 5.19/4.67 & 4.57 (0.04) & 0.28 & 90.81 (0.33) & 0.71 & 4.05/5.14 & 14.52 (0.16) & 0.35 \\ 
 	&    PRR-LAD & 91.75 (0.30) & 0.55 & 4.41/3.84 & 4.95 (0.05) & 0.37 & 91.94 (0.31) & 0.57 & 3.63/4.43 & 15.45 (0.18) & 0.42 \\ 
 	25&    PP & 92.00 (0.31) & 0.56 & 4.21/3.79 & 5.03 (0.05) & 0.40 & 92.47 (0.32) & 0.56 & 3.55/3.98 & 16.31 (0.21) & 0.46 \\ 
 	&  AR-perc & 93.23 (0.28) & 0.42 & 3.59/3.18 & 5.52 (0.07) & 0.49 & 93.41 (0.28) & 0.42 & 2.57/4.02 & 17.36 (0.35) & 0.53 \\ 
 	&    AR-proot & 93.42 (0.29) & 0.45 & 3.59/2.98 & 5.63 (0.07) & 0.51 & 93.24 (0.32) & 0.55 & 2.68/4.08 & 17.75 (0.36) & 0.53 \\ 
 	&    X & 86.87 (0.39) & 1.43 & 6.70/6.43 & 4.28 (0.04) & 0.17 & 87.87 (0.38) & 1.24 & 5.48/6.65 & 13.57 (0.18) & 0.21 \\ 
 	&    QAR-perc & 88.89 (0.37) & 1.04 & 5.74/5.37 & 4.66 (0.06) & 0.26 & 89.84 (0.35) & 0.89 & 4.37/5.79 & 14.91 (0.27) & 0.29 \\ 
	&    QAR-proot & 91.73 (0.33) & 0.64 & 4.44/3.83 & 5.19 (0.07) & 0.42 & 92.37 (0.31) & 0.55 & 2.51/5.12 & 16.64 (0.41) & 0.46 \\ 
  &&&&&&&&&&&\\
			\hline
				&&&&&&&&&&&\\
  	  & ORACLE & 95.00 (0.00) & 0.00 & 2.50/2.50 & 4.78 (0.14) & 0.50 & 95.00 (0.00) & 0.00 & 2.50/2.50 & 14.89 (0.53) & 0.50 \\ 
	  &   BJ & 92.09 (0.19) & 0.27 & 4.04/3.87 & 4.55 (0.03) & 0.28 & 93.05 (0.22) & 0.28 & 1.19/5.75 & 14.47 (0.11) & 0.39 \\ 
 	  &  TS & 92.29 (0.19) & 0.26 & 3.81/3.90 & 4.62 (0.03) & 0.30 & 92.42 (0.23) & 0.34 & 3.90/3.67 & 14.79 (0.13) & 0.34 \\ 
 	  &  CB & 91.99 (0.20) & 0.29 & 3.98/4.03 & 4.58 (0.03) & 0.27 & 91.86 (0.25) & 0.40 & 4.46/3.68 & 14.61 (0.12) & 0.30 \\ 
 	  &  PRR  & 92.50 (0.20) & 0.25 & 3.76/3.74 & 4.65 (0.03) & 0.32 & 92.78 (0.21) & 0.27 & 3.54/3.68 & 14.86 (0.13) & 0.37 \\ 
 	  &  PRR-LAD & 93.20 (0.19) & 0.22 & 3.35/3.44 & 4.83 (0.03) & 0.42 & 93.36 (0.20) & 0.24 & 3.34/3.30 & 15.44 (0.14) & 0.44 \\ 
 	50  &  PP & 93.48 (0.19) & 0.20 & 3.25/3.27 & 4.88 (0.03) & 0.43 & 93.58 (0.23) & 0.30 & 3.31/3.11 & 15.67 (0.14) & 0.45 \\ 
 	  & AR-perc & 94.12 (0.18) & 0.16 & 2.84/3.04 & 5.08 (0.04) & 0.53 & 94.56 (0.18) & 0.16 & 2.46/2.98 & 16.21 (0.14) & 0.55 \\ 
    & AR-proot & 94.16 (0.19) & 0.18 & 2.89/2.95 & 5.10 (0.04) & 0.52 & 94.62 (0.19) & 0.17 & 2.31/3.07 & 16.36 (0.15) & 0.56 \\ 
  	  & X & 91.22 (0.23) & 0.40 & 4.34/4.44 & 4.51 (0.03) & 0.24 & 91.67 (0.22) & 0.35 & 3.79/4.54 & 14.17 (0.12) & 0.28 \\ 
 	  &  QAR-perc & 91.90 (0.22) & 0.34 & 4.04/4.06 & 4.65 (0.03) & 0.31 & 92.44 (0.21) & 0.28 & 3.37/4.19 & 14.62 (0.14) & 0.33 \\ 
 	  &  QAR-proot & 92.92 (0.21) & 0.27 & 3.60/3.47 & 4.84 (0.04) & 0.39 & 93.68 (0.19) & 0.20 & 2.26/4.06 & 15.31 (0.14) & 0.45 \\ 
     &&&&&&&&&&&\\
			\hline
				
   \hline
\end{tabular}}
\caption{Simulation results for Model~1 with $\phi_1=0.6$ and horizon $k=3$, for $n=25$ and $n=50$ with $S=500$. The table reports average coverage $\overline{\hat{\beta}}$ (with its standard error),  mean squared error of the coverage (MSE), tail probabilities $\overline{\hat{B}}$/$\overline{\hat{A}}$, average interval length (with its standard error) and symmetry statistic $\hat{\gamma}$ for  $N(0,1)$ and $\chi^2_5$ innovations. The nominal confidence level is $\beta=0.95$.}
	\label{tab:ar12}
\end{table}
 
The joint analysis of Figure~\ref{fig:ar1} and Tables~\ref{tab:ar1} and~\ref{tab:ar12} reveals that, as the prediction horizon increases, the average coverage provided by the methods tends to decline. This is because the precision in estimating future values decreases with longer horizons, due to the increasing uncertainty introduced by recursively forecasting intermediate steps ($k = 1, \dots, K$) based on observations up to time $n$. Among all methods, the proposed AR-perc, AR-proot, QAR-perc, and QAR-proot, together with method X of \cite{xiao2012}, appear to be the least sensitive to the prediction horizon $k$, both in terms of average coverage and mean squared error. Notably, all these methods are based on quantile techniques.

Table~\ref{tab:ar1} ($k = 1$) also shows that, for symmetric innovation distributions, the methods yield similar average percentages of observations below and above the prediction intervals. However, all methods struggle to capture extremely high future values when innovations follow a positively skewed distribution, such as $\chi^2_5$.
Interestingly, Table~\ref{tab:ar12} reveals that for $k = 3$ and $\chi^2_5$ innovations, the average percentage of uncovered high values is more similar to that of uncovered low values compared to the case of $k = 1$. This could be because the distribution of $Y_{n+k}^{s,f}$, conditional on the series up to time $n$, is the convolution of $k$ distributions obtained by rescaling the innovation distribution, $\chi^2_5$ in this case, 
resulting in a symmetrization effect.  

Regarding interval length, Tables~\ref{tab:ar1} and \ref{tab:ar12} show that, particularly for $k=3$, some methods produce prediction intervals that are even shorter than the ORACLE benchmark, which explains their undercoverage. In contrast, our proposals AR-perc, AR-proot, and QAR-proot consistently yield intervals with lengths exceeding those of the ORACLE. Moreover, as the sample size increases, the lengths of the intervals produced by these methods become closer to the ORACLE benchmark.

\subsubsection{Model 2}
The simulation study includes Model 2 to examine the effect of a larger autoregressive order $p$ in AR($p$) processes on coverage performance. A higher value of 
$p$ implies that more parameters must be estimated to construct the prediction intervals, which can influence their accuracy. To assess this effect, we study the conditional coverage distribution provided by the methods.

\begin{figure}[H]
	\centering
	\includegraphics[width=14cm, keepaspectratio=true]{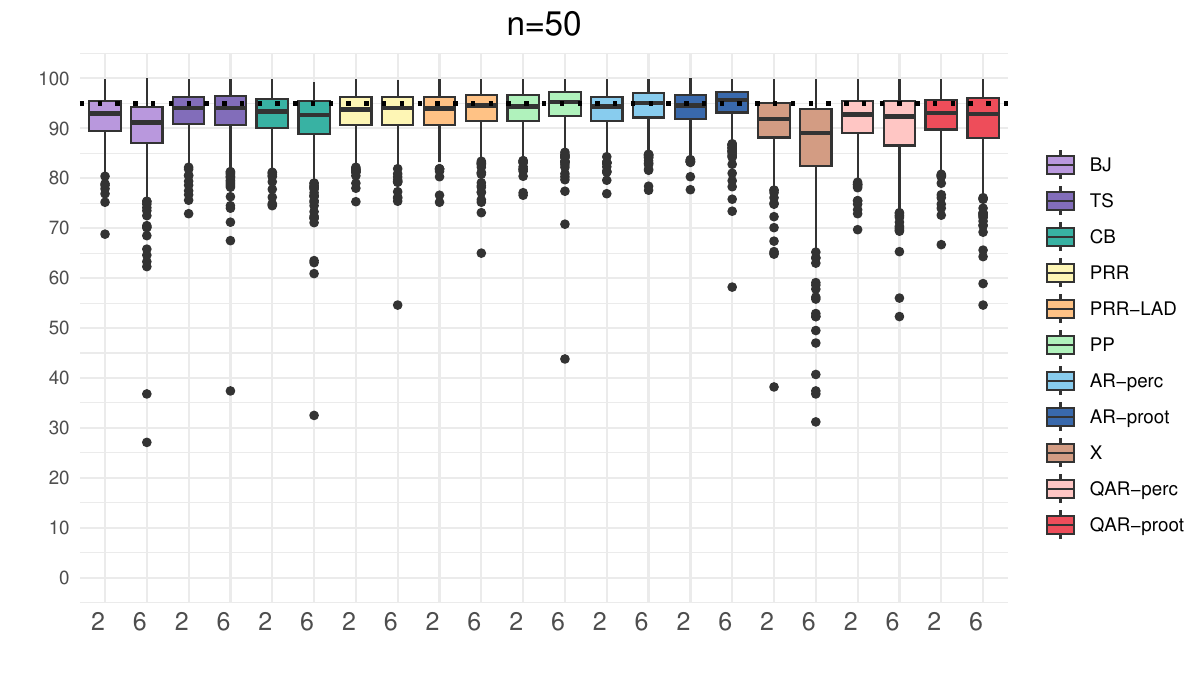}
    \includegraphics[width=14cm, keepaspectratio=true]{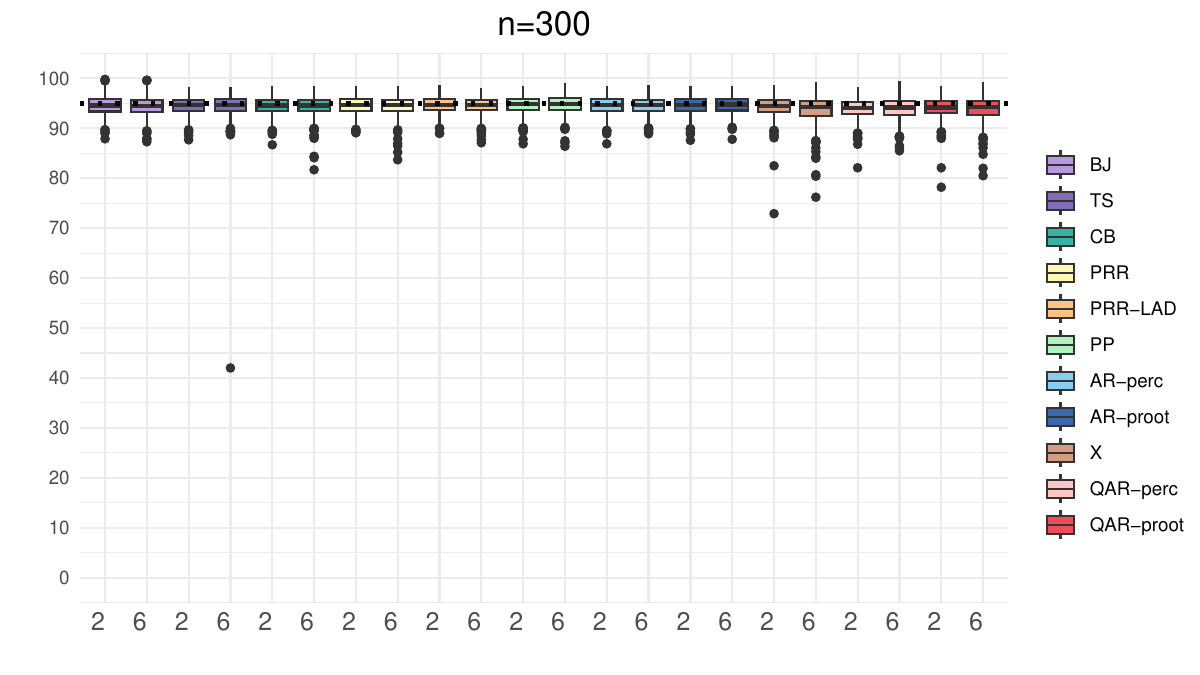}
	\caption{Boxplots of the conditional coverage for each method in Model~2 with $T_3$ innovations and $p=2,6$ (shown on the x-axis). Results are reported for horizon $k=3$ with $n=50$ (top) and $n=300$ (bottom), based on $S=500$ replications. The nominal confidence level is $\beta=0.95$.
} 
	\label{fig:ar2ar6}
\end{figure}

The estimated conditional coverage $\hat{\beta}_s$ (given in (\ref{eq:cov}))  is a random variable that measures the proportion of future values (out of $F=1000$) covered by the method’s prediction interval in each sample, so its distribution across the 
$S=500$ replications can be summarized using a boxplot.
Figure~\ref{fig:ar2ar6} displays the boxplots for the different methods under Model 2, with $p = 2$ and $p = 6$, for $n = 50$ (top) and $n = 300$ (bottom), using $T_3$ innovations, prediction horizon of $k = 3$ and nominal confidence level $\beta=0.95$.  
These boxplots show that the distributions are left-skewed, with some extreme lower outliers, particularly when $p = 6$ and $n = 50$.
The proposed methods, AR-perc and AR-proot, appear to be less affected by the low-coverage issue, maintaining consistent performance from $p = 2$ to $p = 6$ under $T_3$ innovations and $n=50$. In fact, AR-perc is the best-performing percentile-based method (that is, the best among TS, CB, PRR, PRR-LAD, AR-perc, X, QAR-perc), while AR-proot is the most accurate predictive root procedure (that is, the best among PP, AR-proot and QAR-proot) in this scenario.
Figure~\ref{fig:ar2ar6} also provides further evidence that X, QAR-perc, and QAR-proot remain consistent under AR($p$) models, as can be seen by comparing their (conditional) coverage distributions in the top ($n = 50$) and bottom ($n = 300$) panels. 

\subsubsection{Model 3}
Model 3 is a QAR (1). When $U_t > (1-\gamma_0)/\gamma_1$, the model behaves like a unit root model; otherwise, it exhibits mean reversion. 
In the simulation, we used $\gamma_0 = 0.25$ and $\gamma_1 = 0.85$, resulting in a globally stationary process that displays  local unit root behavior with a probability of approximately 0.12. This model enables us to examine the coverage performance of the methods under the QAR($p$) framework in the presence of transient unit root dynamics. Furthermore, we address the key question of whether the methods originally developed to obtain prediction intervals for AR($p$) processes remain consistent in the QAR($p$) context.

\begin{figure}[H]
	\centering
	\includegraphics[width=14cm, keepaspectratio=true]{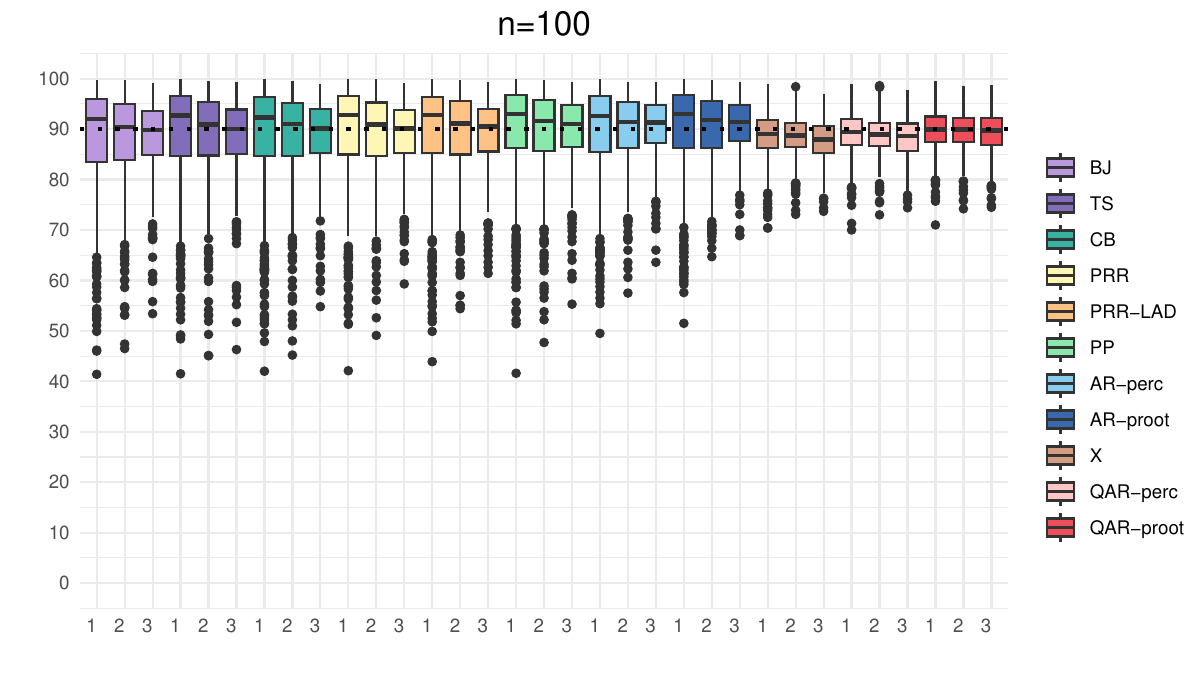}
    \includegraphics[width=14cm, keepaspectratio=true]{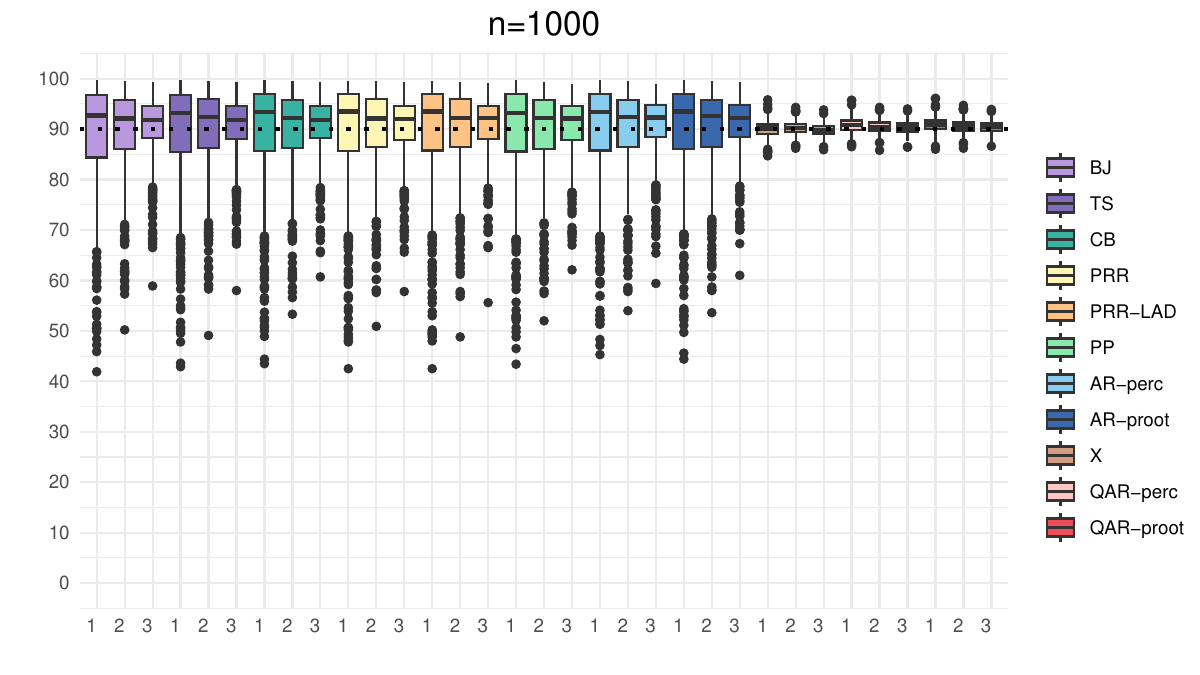}
	\caption{Boxplots of the conditional coverage for each method for Model 3, with $N(0,1)$ distribution for $F_a$ and horizons $k=1,2,3$ (shown on the x-axis).  Results are reported for horizon  $n=100$ (top) and $n=1000$ (bottom), based on $S=500$ replications. The nominal confidence level is $\beta=0.90$.} 
	\label{fig:qar1}
\end{figure}

Figure \ref{fig:qar1} shows the empirical distribution of the (conditional) coverage provided by the methods for Model 3 with $N(0,1)$ distribution for $F_a$ and nominal confidence level $\beta=0.9$, with $n=100$ (top) and $n=1000$ (bottom).
As observed, both existing methods in the literature for AR($p$) process (BJ, TS, CB, PRR, PRR-LAD and PP) and the new methods for AR($p$) process (AR-perc and AR-proot) fail to attain the nominal level, and their performance even deteriorates as the sample size increases, demonstrating clear signs of inconsistency. In contrast,  QAR-specific methods X, QAR-perc, and QAR-proot exhibit improved performance with larger samples, with QAR-proot achieving the best result in terms of median coverage for $n=100$. 
Given these findings, we discard the inconsistent methods from further analysis and concentrate exclusively on the consistent ones.

\subsubsection{Model 4}
The simulation study includes Model 4, a QAR(2) process, to further examine and compare the performance of QAR-specific methods under higher autoregressive orders. 
Table \ref{tab:qar2} presents the simulation results for Model 4 when both $N(0,1)$ and $T_3$ distributions for $F_a$ are considered, with varying prediction horizons ($k=1,3$) and sample sizes ($n=50,100,200$) using nominal confidence level $\beta=0.9$. As expected, the three methods improve average coverage (while also reducing the SE and MSE of the coverage) as the sample size increases. Notably, QAR-proot provides the best overall results in terms of mean coverage, MSE, and $\hat{\gamma}$, followed by QAR-perc as the second-best method. Additionally, the three methods yield prediction intervals with similar average lengths, which are consistent with the average empirical length.

\begin{table}[H]
\centering
\scalebox{0.75}{
\begin{tabular}{|lrr|rrrrr|rrrrr|}
  \cline{4-13}
  \multicolumn{3}{c|}{} & \multicolumn{5}{c}{N(0,1)} & \multicolumn{5}{|c|}{$T_3$} \\
  \hline
$k$ & n & method & $\overline{\hat{\beta}}$ (SE) & MSE & $\overline{\hat{B}}$/$\overline{\hat{A}}$ & $\overline{\textrm{Len}}$(SE) & $\hat{\gamma}$ & $\overline{\hat{\beta}}$ (SE) & MSE & $\overline{\hat{B}}$/$\overline{\hat{A}}$ & $\overline{\textrm{Len}}$(SE) & $\hat{\gamma}$ \\
  \hline

  1 & 50 & ORACLE & 90$\%$ & - & 5.00 / 5.00 & 3.27 (0.77) & 0.50 & 90$\%$& -&5.00 / 5.00& 4.69 (1.25) & 0.50\\
&&X & 85.59 (0.40) & 1.00 & 7.11 / 7.30 & 3.23 (0.04) & 0.35& 86.00 (0.45) & 1.15 & 6.84 / 7.16 & 5.05 (0.09) & 0.40  \\ 
&&  QAR-perc & 86.54 (0.34) & 0.68 & 6.62 / 6.84 & 3.26 (0.04) & 0.35& 87.43 (0.33) & 0.62 & 6.15 / 6.42 & 5.13 (0.09) & 0.42  \\ 
&& QAR-proot & 88.25 (0.31) & 0.51 & 5.70 / 6.05 & 3.43 (0.05) & 0.46& 87.60 (0.26) & 0.40 & 5.75 / 6.65 & 5.86 (0.08) & 0.38 \\ 
 & 100 &  ORACLE & 90$\%$ & - & 5.00 / 5.00 & 3.29 (0.78) & 0.5 & 90$\%$& -&5.00 / 5.00& 4.75 (1.38)& 0.5\\
  &   & X & 87.44 (0.28) & 0.45 & 6.49 / 6.07 & 3.25 (0.04) & 0.36 & 87.95 (0.31) & 0.52 & 6.14 / 5.91 & 4.90 (0.07) & 0.40 \\
  &   & QAR-perc & 87.64 (0.24) & 0.34 & 6.31 / 6.04 & 3.23 (0.04) & 0.33 & 88.51 (0.20) & 0.23 & 5.79 / 5.70 & 4.87 (0.07) & 0.40 \\
  &   & QAR-proot & 88.47 (0.23) & 0.29 & 5.85 / 5.68 & 3.31 (0.04) & 0.41 & 88.51 (0.22) & 0.26 & 5.74 / 5.75 & 4.91 (0.07) & 0.41 \\
  & 200 &  ORACLE & 90$\%$ & - & 5.00 / 5.00 & 3.32 (0.81) & 0.5 & 90$\%$& -&5.00 / 5.00& 4.75 (1.31)& 0.5\\
  &   & X & 88.56 (0.21) & 0.23 & 5.67 / 5.77 & 3.29 (0.04) & 0.38 & 88.88 (0.23) & 0.29 & 5.58 / 5.53 & 4.87 (0.06) & 0.45 \\
  &   & QAR-perc & 88.96 (0.17) & 0.16 & 5.60 / 5.44 & 3.31 (0.04) & 0.42 & 89.47 (0.18) & 0.16 & 5.34 / 5.18 & 4.93 (0.06) & 0.50 \\
  &   & QAR-proot & 89.47 (0.17) & 0.15 & 5.34 / 5.20 & 3.37 (0.04) & 0.48 & 89.65 (0.18) & 0.15 & 5.28 / 5.07 & 4.99 (0.06) & 0.52 \\

  \hline
3& 50 &  ORACLE & 90$\%$ & - & 5.00 / 5.00 &  3.76 (0.57)  & 0.5 & 90$\%$ & - & 5.00 / 5.00& 5.60 (0.91) & 0.50 \\
& & X& 85.52 (0.29) & 0.62 & 6.89 / 7.59 & 3.54 (0.03) & 0.27& 85.65 (0.30) & 0.64 & 6.92 / 7.43 & 5.51 (0.07) & 0.29 \\ 
  & & QAR-perc& 86.94 (0.27) & 0.46 & 6.35 / 6.71 & 3.69 (0.04) & 0.36 & 86.83 (0.28) & 0.48 & 6.45 / 6.71 & 5.73 (0.08) & 0.35 \\ 
& & QAR-proot & 88.48 (0.26) & 0.37 & 5.05 / 6.47 & 3.88 (0.04) & 0.45 &  87.60 (0.26) & 0.40 & 5.75 / 6.65 & 5.86 (0.08) & 0.38 \\ 
 & 100 & EMP & 90$\%$ & - & 5.00 / 5.00 & 3.78 (0.60) & 0.5 & 90$\%$ & - & 5.00 / 5.00& 5.66 (1.06) & 0.5 \\
  &   & X & 87.63 (0.18) & 0.22 & 5.91 / 6.46 & 3.66 (0.03) & 0.30 & 87.60 (0.20) & 0.25 & 5.95 / 6.45 & 5.61 (0.07) & 0.31 \\
  &   & QAR-perc & 88.36 (0.17) & 0.17 & 5.80 / 5.84 & 3.75 (0.03) & 0.36 & 88.23 (0.19) & 0.21 & 5.85 / 5.92 & 5.75 (0.08) & 0.35 \\
  &   & QAR-proot & 89.25 (0.17) & 0.16 & 5.05 / 5.70 & 3.85 (0.04) & 0.45 & 88.67 (0.19) & 0.19 & 5.50 / 5.83 & 5.84 (0.08) & 0.39 \\
  & 200 &  ORACLE & 90$\%$ & - & 5.00 / 5.00 & 3.82 (0.60) & 0.5 & 90$\%$ & - & 5.00 / 5.00 & 5.68 (1.02) & 0.5 \\
  &   & X & 88.82 (0.13) & 0.10 & 5.30 / 5.88 & 3.75 (0.03) & 0.37 & 88.72 (0.17) & 0.15 & 5.35 / 5.93 & 5.63 (0.05) & 0.37 \\
  &   & QAR-perc & 89.00 (0.12) & 0.09 & 5.43 / 5.56 & 3.77 (0.03) & 0.38 & 89.00 (0.15) & 0.12 & 5.45 / 5.55 & 5.69 (0.05) & 0.40 \\
  &   & QAR-proot & 89.43 (0.12) & 0.08 & 5.12 / 5.45 & 3.82 (0.03) & 0.43 & 89.10 (0.14) & 0.11 & 5.35 / 5.54 & 5.71 (0.05) & 0.40 \\
  \hline
\end{tabular}}
\caption{Simulation results for Model~4 with horizons $k=1$ and $k=3$, for $n=50,100,$ and $200$ with $S=500$. The table reports the average conditional coverage $\overline{\hat{\beta}}$ (with its standard error), mean squared error (MSE), upper and lower tail probabilities $\overline{\hat{A}}/\overline{\hat{B}}$, average interval length (with standard error), and the symmetry statistic $\hat{\gamma}$ for  $N(0,1)$ and $T_3$ distributions for $F_a$. The nominal confidence level is $\beta=0.90$.
}
\label{tab:qar2}
\end{table}
\subsection{Computational time}

Finally,
Table \ref{tab:time} reports the average execution times (in seconds) over $S=500$ replications of Model 1 with $\phi_1=0.6$, N(0,1) innovations' distribution and $k=4$, required for the methods to produce the prediction intervals. As before, for the AR($p$)-based methods we considered $B=1000$ bootstrap replications, whereas for the QAR($p$)-based procedures we used $B=5000$ bootstrap replications. When needed, the quantile regression estimates were obtained using the \texttt{quantreg} package.

\begin{table}[H]
\centering
\scalebox{0.85}{
\begin{tabular}{|c|rrrrrrrr|rrr|}
\hline
 \multicolumn{1}{|c|}{}& \multicolumn{8}{c}{AR($p$)-based methods} & \multicolumn{3}{|c|}{QAR($p$)-based methods} \\
\cline{2-12}
  $n$   & BJ & TS & CB & PRR & PRR-LAD & PP & AR-perc & AR-proot & X & QAR-perc & QAR-proot \\
\hline
50  & 0.11 & 0.88 & 0.04 & 0.90 & 1.09 & 1.49 & 0.83 & 0.93 & 14.34 & 15.60 & 18.49 \\
100 & 0.13 & 1.31 & 0.04 & 1.33 & 1.56 & 2.10 & 0.95 & 1.14 & 16.48 & 17.84 & 21.22 \\
200 & 0.13 & 1.88 & 0.04 & 1.94 & 2.19 & 2.86 & 0.98 & 1.34 & 16.87 & 18.30 & 21.73 \\
500 & 0.17 & 3.86 & 0.05 & 4.07 & 4.40 & 5.59 & 1.18 & 2.12 & 20.76 & 22.74 & 26.77 \\
\hline
\end{tabular}}
\caption{Average execution times (in seconds) for different methods with $k = 4$ and varying $n$, based on $S=500$. $B=1000$ was considered for AR($p$)-based methods, and $B=5000$ for QAR($p$)-based methods.}
\label{tab:time}
\end{table}

Focusing on bootstrap-based procedures, and excluding the CB method, which does not account for the variability induced by parameter estimation, we observe that AR-perc is the most computationally efficient among the percentile-based approaches (TS, CB, PRR, PRR-LAD, AR-perc, X and QAR-perc). Likewise, within the predictive root approach (PP, AR-proot and QAR-proot), AR-proot attains the shortest average execution time. The efficiency of both AR-perc and AR-proot stems from the fact that they circumvent the resampling of the original time series and incorporate parameter uncertainty directly through a multiplier bootstrap scheme.
By comparing percentile-based and predictive root-based methods for the AR($p$) model, we observe that the former are computationally more efficient, as common residuals are used in percentile-based methods, whereas predictive residuals are used in predictive root-based methods. Regarding prediction interval approaches for QAR($p$), the higher number of bootstrap replications accounts for the greater computational cost. Similar to CB, Method X does not account for the variability induced by parameter estimation, which reduces its computational cost.

\section{Empirical applications} 
Two real economic time series illustrate the usage of the proposed bootstrap procedures AR-perc, AR-proot, QAR-perc and QAR-proot. As in the simulation, we compare their performance with already-existing methods.

\subsection{United States unemployment rate}
\label{sec:app1}

 Several studies suggest that unemployment responds asymmetrically to shocks, surging quickly in recessions but easing slowly in expansions. This behavior could give rise to a heteroskedastic time series. 
 Motivated by this possibility, we  consider the U.S. unemployment series in this real data analysis. The data consists of the semiannual unemployment rate (seasonally adjusted) from the first half of 1948 to the first half of 2025, for a total of 155 observations. These data can be recovered directly from \href{https://fred.stlouisfed.org/series/UNRATE}{the web page of St Louis Bank}.

Preliminary autocorrelation analysis suggests that the underlying dependence structure of the series can be adequately captured by an autoregressive specification. Both the Bayesian Information Criterion (BIC) by \cite{schwarz1978} and the partial autocorrelation function select $p=2$ as the lag order for the series.
We use the Augmented Dickey-Fuller (ADF) test \citep{dickey1979,said1984} to check the stationarity assumption, whose result suggests that the series is stationary.

To compare the performance of the prediction intervals, we adopt the rolling-window pseudo-out-of-sample (rwPOOS) procedure of \cite{wu2024}. A training window of $R$ observations $(Y_1,\dots,Y_R)$ is used to forecast $Y_{R+k}$. The window is then moved forward by one observation to $(Y_2,\dots,Y_{R+1})$ to forecast $Y_{R+1+k}$, and so on, until the end of the available dataset. At the end, for each $k$, we compute the proportion of pseudo out-of-sample observations contained in the prediction intervals produced by a method and the average length of these intervals.
We set the training window to $R=50$  to ensure a sufficient number of pseudo out-of-sample forecasts, evaluate $k$-step-ahead horizons for $k=1,2,3,4$, and use a nominal confidence level of $0.95$. Following the simulation study, we employ \(B=1000\) bootstrap replications for the AR(\(p\))-based methods and \(B=5000\) for the QAR(\(p\))-based methods to obtain the prediction intervals. The results, reported in Table~\ref{tab:app1}, summarize (i) the estimated coverage $\hat{\beta}_k$ at each prediction horizon (in percent); (ii) the average absolute deviation from the nominal confidence level, $\overline{D}=\frac{1}{4}\sum_{k=1}^{4}\lvert \hat{\beta}_k - \beta \rvert$, computed across the four horizons (in percent); and (iii) the average length of the prediction intervals for each $k$, $\overline{\textrm{len}}_k$.

Table~\ref{tab:app1} shows that several methods exhibit satisfactory performance for $k=1$. In this case, AR-proot achieves coverage closest to the nominal level, followed by QAR-proot and PRR. As the prediction horizon increases, overall performance deteriorates. Nevertheless, AR-proot remains the best-performing method for $k=2$, followed by QAR-proot and AR-perc. For $k=3$, PP attains the most accurate coverage, with AR-perc ranking second, while for $k=4$ AR-perc provides the best performance. 
Considering all horizons jointly, the aggregate measure reported in column $\overline{D}$ indicates that AR-perc delivers the best overall performance, followed by AR-proot and QAR-proot. In contrast, method X consistently yields the poorest coverage across all horizons ($k=1,2,3,4$).

\begin{table}[H]
\centering
\scalebox{0.85}{
\begin{tabular}{|l|rrrr|c|rrrr|}
\cline{2-5} 
\cline{6-10}
\multicolumn{1}{c|}{}&\multicolumn{4}{|c|}{$\hat{\beta}_k$ }&\multicolumn{1}{c}{$\overline{D}$} &\multicolumn{4}{|c|}{$\overline{\textrm{len}}_k$} \\
\hline
k & 1 & 2 & 3 & 4& - & 1 & 2 & 3 & 4 \\
\hline
BJ        & 93.14 & 86.27 & 83.33 & 82.35 &8.73& 1.89 & 3.26 & 4.26 & 4.93 \\
TS        & 92.16 & 87.25 & 88.24 & 85.29 & 6.76 &2.16 & 3.66 & 4.72 & 5.45 \\
CB        & 92.16 & 87.25 & 88.24 & 81.37 & 7.75&2.14 & 3.61 & 4.51 & 5.10 \\
PRR       & 94.12 & 88.24 & 86.27 & 82.35 & 7.25&2.24 & 3.71 & 4.72 & 5.41 \\
PRR-LAD   & 93.14 & 90.20 & 88.24 & 86.27 & 5.54  &2.29 & 3.75 & 4.75 & 5.39 \\
PP      & 93.14 & 89.22 & 91.18 & 88.24 & 4.56&2.49 & 4.36 & 5.56 & 6.33 \\
AR-perc   & 92.16 & 91.18 & 90.20 & 92.16 & 3.58 &2.31 & 4.10 & 5.73 & 7.23 \\
AR-proot  & 95.10 & 92.16 & 89.22 & 88.24 &3.87 &2.46 & 4.30 & 5.76 & 7.02 \\
X    & 88.24 & 79.41 & 74.51 & 76.47 & 15.34&1.93 & 3.15 & 4.15 & 4.85 \\
QAR-perc  & 91.18 & 90.20 & 82.35 & 81.37 &  8.73&1.87 & 3.52 & 4.95 & 6.21 \\
QAR-proot & 94.12 & 91.18 & 88.24 & 89.22 & 4.31&2.08 & 3.79 & 5.34 & 6.91 \\
\hline
\end{tabular}
}
\caption{Estimated coverage for prediction horizons $k=1,2,3,4$ (in percent), average absolute deviation from the nominal confidence level (in percent), and average interval length for the U.S. unemployment rate series. Results are based on $R=50$. The nominal confidence level is $\beta=0.95$.}
\label{tab:app1}
\end{table}

If we compare interval length in Table \ref{tab:app1}, we see that predictive root method PP delivers the longest intervals for $k=1$ and $k=2$, followed by AR-proot. For horizons $k=3$ and $k=4$ the longest intervals are provided by AR-perc and AR-proot.

\subsection{United States retail gasoline price}

\cite{koenker2006} studied asymmetric dynamics in U.S. retail gasoline prices. 
The dataset they considered consists of weekly data on regular gasoline retail prices from August 20, 1990, to February 16, 2004, for a total of 699 observations. 
The evidence they collected suggested that the series displays local stationary behavior at lower quantiles and local unit-root behavior at higher quantiles. 
They formally tested the null hypothesis of (global) unit-root behavior obtaining a rejection, and also tested for asymmetric dynamics, concluding that the QAR framework better characterizes the series.

We consider the same series as \cite{koenker2006} and use $p=4$, as they suggested. To compare the performance of the prediction intervals, we again adopt the rwPOOS procedure. The availability of data allows us to use a training window of $R=600$ observations. We obtain prediction intervals for $k$-step-ahead horizons, with $k=1,2,3,4$, using a nominal confidence level of $0.95$ and the same settings as in the previous real-data application (Section~\ref{sec:app1}).

\begin{table}[H]
\centering
\scalebox{0.8}{
\begin{tabular}{|l|rrrr|c|rrrr|}
\cline{2-5} 
\cline{6-10}
\multicolumn{1}{c|}{} & \multicolumn{4}{c|}{$\hat{\beta}_k$} & \multicolumn{1}{c|}{$\overline{D}$} & \multicolumn{4}{c|}{$\overline{\textrm{len}}_k$} \\
\hline
k & 1 & 2 & 3 & 4 & - & 1 & 2 & 3 & 4 \\
\hline
BJ        & 79.17 & 82.29 & 82.29 & 83.33 & 13.23 & 0.06 & 0.11 & 0.16 & 0.21 \\
TS        & 84.38 & 88.54 & 87.50 & 87.50 &  8.02 & 0.08 & 0.13 & 0.18 & 0.22 \\
CB        & 85.42 & 85.42 & 84.38 & 87.50 &  9.32 & 0.07 & 0.12 & 0.17 & 0.22 \\
PRR       & 83.33 & 84.38 & 83.33 & 85.42 & 10.89 & 0.07 & 0.12 & 0.17 & 0.21 \\
PRR-LAD   & 85.42 & 84.38 & 83.33 & 86.46 & 10.10 & 0.07 & 0.12 & 0.17 & 0.22 \\
PP     & 87.50 & 87.50 & 85.42 & 87.50 &  8.02 & 0.07 & 0.13 & 0.18 & 0.22 \\
AR-perc   & 86.46 & 86.46 & 85.42 & 88.54 &  8.28 & 0.07 & 0.12 & 0.18 & 0.23 \\
AR-proot  & 85.42 & 87.50 & 85.42 & 88.54 &  8.28 & 0.07 & 0.13 & 0.18 & 0.23 \\
X     & 96.88 & 95.83 & 97.92 & 93.75 &  1.72 & 0.09 & 0.17 & 0.24 & 0.30 \\
QAR-perc  & 96.88 & 95.83 & 97.92 & 95.83 &  1.61 & 0.09 & 0.17 & 0.24 & 0.31 \\
QAR-proot & 96.88 & 96.88 & 97.92 & 95.83 &  1.88 & 0.10 & 0.17 & 0.25 & 0.32 \\
\hline
\end{tabular}
}
\caption{Estimated coverage for prediction horizons $k=1,2,3,4$ (in percent), average absolute deviation from the nominal confidence level (in percent), and average interval length for the U.S.\ retail gasoline price series. Results are based on $R=600$. The nominal confidence level is $\beta=0.95$.}
\label{tab:app2}
\end{table}

Table~\ref{tab:app2} reports the estimated coverage at each prediction horizon, the average absolute deviation from the nominal confidence level, and the average lengths of the prediction intervals.
As we can observe, the AR($p$)-based methods show poor performance, with coverages below $0.90$ despite the nominal confidence level being $0.95$. The QAR($p$)-based methods perform much better offering similar results, with QAR-perc providing the best overall results, as indicates column $\overline{D}$.

\begin{figure}[H]
\centering
\includegraphics[width=12cm, keepaspectratio=true]{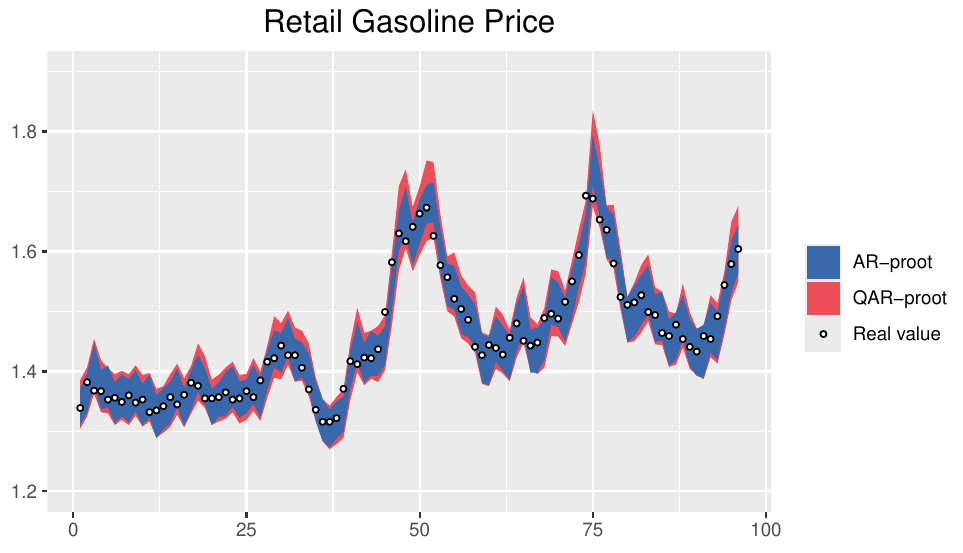}
\caption{Prediction intervals obtained by AR-proot and QAR-proot methods for the retail gasoline price series for $k=1$.}
\label{fig:app2}
\end{figure}

Regarding interval length, in this case, the QAR($p$)-based methods exhibit similar results to each other but produce longer intervals than the AR($p$)-based methods. This bigger length seems to play an important role to attain consistent coverage. Figure \ref{fig:app2}, that compares the intervals produced by QAR-proot and AR-proot methods for $k=1$, supports this affirmation.

\section{Conclusions}

In this work, prediction intervals for the classical autoregressive model are reviewed, and new methods are proposed to combine components that provide improved performance from both statistical and computational perspectives. In particular, quantile estimation of coefficients and bootstrap multipliers have been shown to be useful in achieving generally competitive coverage with low computational cost. One proposal, called AR-perc, is a percentile-based method, while the other, called AR-proot, is based on replicating a predictive root. The validity and pertinence of the proposed methods are established, and a detailed simulation study demonstrates the advantages of the new proposals.

Within the framework of the quantile autoregressive model, QAR($p$), to the best of our knowledge, only the proposal by \cite{xiao2012} was previously available in the literature. However, this method does not account for the randomness arising from coefficient estimation. Therefore, two new methods, QAR-perc and QAR-proot, are proposed, following percentile-based and predictive-root-based approaches, respectively. The validity and pertinence of each method are established. Simulation results show that the new methods outperform the existing proposal by \cite{xiao2012}.

The general conclusion drawn from the simulation study is that the new methods improve upon their natural competitors. In the context of the classical autoregressive model, the AR-perc method is preferable among percentile-based methods, while AR-proot is preferable among predictive-root-based methods. In the context of the quantile autoregressive model, only methods specifically designed for this framework appear to be consistent, and the new proposals exhibit good overall performance. Moreover, although the new methods designed for the more general QAR($p$) model are less efficient under the classical AR($p$) setting, they remain consistent and provide acceptable performance.

The new methods are also illustrated using real data. For the unemployment series, despite indications of heteroskedasticity, the proposed AR($p$)-based methods remain accurate for constructing prediction intervals. In the case of retail gasoline prices, the asymmetric dynamics and local unit-root behavior identified by \cite{koenker2006} appear to deteriorate the performance of AR($p$)-based methods. The strong performance of the QAR($p$)-based methods further supports the conclusion that the QAR($p$) framework is more appropriate for this series.

Overall, the new QAR-proot method can be recommended as a general choice for constructing prediction intervals, as it adapts well to the general QAR($p$) model while remaining reasonably efficient under the more restrictive AR($p$) model.

\appendix

\section*{Appendix}

{\it Proof of Lemma \ref{l:AR-perc}} To prove part (a), we will consider the random function
$$R_n(v)=\frac{1}{n-p}\sum_{t=p+1}^n\left[\rho_\tau\left(a_t-v^tZ_{t,p}\right)-\rho_\tau\left(a_t\right)\right],$$
where $v\in \mathbb{R}^{p+1}$. Note that the minimizer of $R_n(v)$, say $\hat{v}$, will be $\hat{v}=\hat{\phi}-\phi$. Then, the consistency of $\hat{ \phi}$ will be equivalent to the convergence of $\hat{v}$ to zero.

Using Knight's identity (see \citealp{knight1998}), for $u\neq 0$,
$$\rho_\tau\left(u-v\right)-\rho_\tau\left(u\right)=-v\,\psi_\tau(u)+\int_0^v \left[I(u\leq s)-I(u<0)\right]\, ds,$$
where $\psi(u)=\tau-I(u<0)$, so we have the decomposition
$$R_n(v)=-\frac{1}{n-p}\sum_{t=p+1}^n v^tZ_{t,p} \psi_\tau\left(a_t\right)+\frac{1}{n-p}\sum_{t=p+1}^n\int_0^{v^tZ_{t,p}} \left[I(a_t\leq s)-I(a_t<0)\right]\, ds\equiv -A_n(v)+B_n(v).$$

First, we will deal with $A_n(v)$. If ${\cal F}_t$ denotes the $\sigma$-field generated by $\left\{y_s, s\leq t\right\}$, then
$$E\left[v^tZ_{t,p} \psi_\tau\left(a_t\right)| {\cal F}_{t-1}\right]=v^tZ_{t,p} E\left[\psi_\tau\left(a_t\right)\right]=0,$$
where it was used that $Z_{t,p}$ is ${\cal F}_{t-1}$-measurable and $a_t$ is independent of ${\cal F}_{t-1}$ and has $\tau$-quantile zero. Then, the summands of $A_n(v)$, $\left\{v^tZ_{t,p} \psi_\tau\left(a_t\right)\right\}$, are a martingale difference sequence, so $A_n(v)$ will converge to zero almost surely as soon as we prove the condition required by \cite{chow1967} for the strong law of large numbers for martingales, that is,
$$\sum_{t=p+1}^n \frac{E\left[v^tZ_{t,p} \psi_\tau\left(a_t\right)\right]^2}{(t-p)^2}=\tau(1-\tau)\, v^t \Sigma_1 v \sum_{t=p+1}^n\frac{1}{(t-p)^2}<\infty,$$
where $\Sigma_1=E\left[Z_{t,p}\,Z_{t,p}^t\right]$ does not depend on $t$ thanks to the stationarity of the AR($p$) process.

With respect to $B_n(v)$, we denote
$B_n(v)=\frac{1}{n-p}\sum_{t=p+1}^n \xi_t(v)$, where
$$\xi_t(v)=\int_0^{v^tZ_{t,p}} \left[I(a_t\leq s)-I(a_t<0)\right]\, ds.$$
Now, we consider the conditional expectations, $\bar{\xi}_t(v)=E\left[\xi_t(v)|{\cal F}_{t-1}\right]$, and consequently $\bar{B}_n(v)=\frac{1}{n-p}\sum_{t=p+1}^n \bar{\xi}_t$. Then $\left\{\xi_t(v)-\bar{\xi}_t(v)\right\}$
is a martingale difference sequence, so $B_n(v)-\bar{B}_n(v)$ will converge to zero almost surely, since
$$\sum_{t=p+1}^n \frac{E\left|v^tZ_{t,p}\right|^2}{(t-p)^2}=v^t\Sigma_1 v\sum_{t=p+1}^n \frac{1}{(t-p)^2}<\infty.$$

While $A_n(v)$ and $(B_n(v)-\bar{B}_n(v))$ were proven to converge to zero, $\bar{B}_n(v)$ will be bounded away from zero. To see this, we consider
\begin{eqnarray*}
    \bar{\xi}_{t,p}(v)&=&E\left[\left.\int_0^{v^tZ_{t,p}} \left[I(a_t\leq s)-I(a_t<0)\right]\, ds\right|{\cal F}_{t-1}\right] \\
    &=&\int_0^{v^tZ_{t,p}} \left[F_a(s)-F_a(0)\right]\, ds\geq m\int_0^{v^tZ_{t,p}} s\, ds=\frac{m}{2}v^tZ_{t,p}Z_{t,p}^tv,    
\end{eqnarray*}
where the inequality is obtained by a mean value theorem and the boundedness of $f_a$ away from zero. Then, we have the following lower bound for $\bar{B}_n(v)$:
$$\bar{B}_n(v)\geq \frac{m}{2} \frac{1}{n-p}\sum_{t=p+1}^nv^tZ_{t,p}Z_{t,p}^tv.$$
This lower bound is a random convex function that converges to $(m/2)v^t\Sigma_1 v$ almost surely, for all $v$. Then, by Lemma 2.1 in \cite{raozhao1992}, this convergence is uniform. Note also that since $\Sigma_1$ is positive definite, $v^t\Sigma_1v$ is bounded away from zero, that is, for any $\Delta>0$, $\inf_{\|v\|=\Delta}v^t\Sigma_1v>0$. It can then be concluded that
$$\liminf_{n\rightarrow\infty} \inf_{\|v\|=\Delta} \bar{B}_n(v)>0\qquad\text{almost surely.}$$

To finish the proof, we use the argument in the proof of Theorem 2.1 in \cite{Bassett_Koenker_1986}. Thus, we observe that $R_n(0)=0$, so $R_n(v)\leq 0$ in its minimum set. Then, since $R_n(v)$ is a convex function, the convergence of the minimizer of $R_n(v)$, $\hat v$, to zero almost surely will be shown if we can prove that for any $\Delta>0$,
\begin{equation} \label{eq:inf}
    \liminf_{n\rightarrow\infty} \inf_{\|v\|=\Delta} R_n(v)>0\qquad\text{almost surely.}    
\end{equation}
This fact was already obtained for $\bar{B}_n(v)$, so to prove (\ref{eq:inf}) it only remains to show that the convergence of $A_n(v)$ and $(B_n(v)-\bar{B}_n(v))$ to zero is uniform. To do this, we observe that
$$\left|A_n(v)-A_n(v_0)\right|=\left|\frac{1}{n-p}\sum_{t=p+1}^n (v-v_0)^tZ_{t,p} \psi_\tau\left(a_t\right)\right|\leq \left\|v-v_0\right\|\ \frac{1}{n-p}\sum_{t=p+1}^n \left\|Z_{t,p}\right\|,$$
where $\frac{1}{n-p}\sum_{t=p+1}^n \left\|Z_{t,p}\right\|$ converges almost surely to $E\|Z_{t,p}\|$ as $n$ goes to infinity.

Similarly,
\begin{eqnarray*}
    \left|B_n(v)-\bar{B}_n(v)-\left[B_n(v_0)-\bar{B}_n(v_0)\right]\right| \qquad\qquad\qquad\qquad\qquad\qquad\qquad\qquad\qquad \\
      =\left|\frac{1}{n-p}\sum_{t=p+1}^n  \int_0^{v^tZ_{t,p}} \left[I(a_t\leq s)-I(a_t<0)-\left(F_a(s)-F_a(0)\right)\right]\, ds\right|  \\
     \leq \frac{1}{n-p}\sum_{t=p+1}^n \left|v^tZ_{t,p}-v_0^tZ_{t,p}\right\|\leq \left\|v-v_0\right\|\ \frac{1}{n-p}\sum_{t=p+1}^n \left\|Z_{t,p}\right\|,
\end{eqnarray*}
which completes the proof of uniform convergence to zero and then of part (a).

To prove part (b), we make use of the following decomposition, already considered in \cite{boldin1983}:
\begin{eqnarray*}
    \hat{F}_n(x)-F_a(x)&=&\frac{1}{n-p}\sum_{t=p+1}^n \left[I\left(\hat{a}_t\leq x\right)-F_a(x)\right] \\
    &=&\frac{1}{n-p}\sum_{t=p+1}^n\left[I\left(a_t\leq x+(\hat{\phi}-\phi)^t Z_{t,p}\right)-F_a(x)\right] = W_{n1}(x)+W_{n2}(x),
\end{eqnarray*}
where
\begin{eqnarray*}
    W_{n1}(x) &=& \frac{1}{n-p}\sum_{t=p+1}^n \left[I\left(a_t\leq x+(\hat{\phi}-\phi)^t Z_{t,p}\right)-F_a\left(x+(\hat{\phi}-\phi)^t Z_{t,p}\right)\right], \\
    W_{n2}(x) &=& \frac{1}{n-p}\sum_{t=p+1}^n \left[F_a\left(x+(\hat{\phi}-\phi)^t Z_{t,p}\right)-F_a(x)\right].
\end{eqnarray*}

With respect to $W_{n2}(x)$, it can be bounded by
\begin{eqnarray*}
    \sup_x\left|W_{n2}(x)\right|&\leq& \sup_x\frac{1}{n-p}\sum_{t=p+1}^n \left|F_a\left(x+(\hat{\phi}-\phi)^t Z_{t,p}\right)-F_a(x)\right| \\
    &\leq& \sup_x\left|f_a(x)\right| \left\|\hat{\phi}-\phi\right\| \ \frac{1}{n-p}\sum_{t=p+1}^n\left\|Z_{t,p}\right\|,
\end{eqnarray*}
which converges to zero almost surely, due to the convergence of $\hat{\phi}$ to $\phi$ and the boundedness of the other two factors.

To deal with $W_{n1}(x)$, we consider the following uniform bound.
$$\sup_x\left|W_{n1}(x)\right|\leq \sup_{x,\omega} \left|
\frac{1}{n-p}\sum_{t=p+1}^n \left[I\left(a_t\leq x+\omega^t Z_{t,p}\right)-F_a\left(x+\omega^t Z_{t,p}\right)\right] \right|.$$

The supremum on the right side converges to zero almost surely, due to the uniform strong law for martingale processes. To prove this, we first define
$$D_t(x,\omega)=I\left(a_t\leq x+\omega^t Z_{t,p}\right)-F_a\left(x+\omega^t Z_{t,p}\right).$$
Then, for a fixed $(x,\omega)\in\mathbb{R}\times\mathbb{R}^{p+1}$, $\{D_t(x,\omega)\}$ is a martingale difference sequence and satisfies $\left|D_t(x,\omega)\right|\leq 1$. Then, by Lemma 3.1 in \cite{cheng2015} (see also \citealp{azuma1967}), we have that for any $\varepsilon>0$,
$$P\left(\left|\sum_{t=p+1}^n D_t(x,\omega)\right|>\varepsilon\right)\leq 2e^{-\frac{\varepsilon^2}{2(n-p)}}.$$
This exponential inequality, together with the fact that $\{(a,z)\rightarrow I\left(a\leq x+\omega^t z\right): (x,\omega)\in\mathbb{R}\times\mathbb{R}^{p+1}\}$ is a VC-class of functions bounded by one, leads to the conclusion that they satisfy a uniform strong law of large numbers, that is, $\sup_{x,\omega}\left|\frac{1}{n-p}\sum_{t=p+1}^n D_t(x,\omega)\right|$ converges almost surely to zero. We refer to \cite{vandervaart1996} for a detailed study of uniform strong laws with respect to certain classes of functions.

\hfill $\Box$

\section*{Acknowledgements}

The first author acknowledges funding from Universidad Carlos III de Madrid through the Grants for Young PhDs program of the University’s Own Research and Transfer Program 2024. Both the first and second authors were supported by Grant PID2020-116587GB-I00, funded by MCIN/AEI/10.13039/
501100011033. The second author was also supported by the Competitive Reference Groups 2021--2024 program (ED431C 2021/24), funded by the Xunta de Galicia.

\bibliography{BiblioQAR}

\end{document}